\documentclass{agujournal2018}
\usepackage{apacite}
\usepackage{url} 
\usepackage{lineno}

\usepackage{hyperref}
\usepackage{makecell}
\usepackage{arydshln, collcell}
\usepackage{subcaption}
\usepackage{colortbl}
\usepackage{array}
\usepackage{textcomp}
\usepackage[flushleft]{threeparttable}

\draftfalse

\newcommand{\aap}{    {\it Astron. Astrophys. }}

\newcommand{\apjl}{   {\it Astrophys. J. Lett. }}
\newcommand{\apjs}{   {\it Astrophys. J. Suppl. S.}}

\graphicspath{{./}{figures/}}

\journalname{Space Weather}

\begin{document}

%
%


\title{Drag-based CME modeling with heliospheric images incorporating frontal deformation: ELEvoHI 2.0}

%
%




\authors{J\"urgen Hinterreiter\affil{1,2}, Tanja Amerstorfer\affil{1}, Manuela Temmer\affil{2}, Martin~A.~Reiss\affil{1}, Andreas J. Weiss\affil{1,2}, Christian M\"ostl\affil{1}, Luke A. Barnard\affil{3}, Jens Pomoell\affil{4}, Maike Bauer\affil{1,2}, Ute V. Amerstorfer\affil{1}}
\affiliation{1}{Space Research Institute, Austrian Academy of Sciences, Schmiedlstraße 6, 8042 Graz, Austria}
\affiliation{2}{University of Graz, Institute of Physics, Universitätsplatz 5, 8010 Graz, Austria}
\affiliation{3}{Department of Meteorology, University of Reading, Reading, UK}
\affiliation{4}{University of Helsinki, 00100 Helsinki, Finland}





\correspondingauthor{J\"urgen Hinterreiter}{juergen.hinterreiter@oeaw.ac.at}

\begin{keypoints}
\item The implementation of a deformable front based on ELEvoHI for three different ambient solar winds models is presented
\item The parameters influencing the propagation of the CME are studied in detail
\item An estimate of the CME mass is obtained depending on DBM fitting and the cross-sectional area of the CME
\end{keypoints}

\begin{abstract}
The evolution and propagation of coronal mass ejections (CMEs) in interplanetary space is still not well understood. As a consequence, accurate arrival time and arrival speed forecasts are an unsolved problem in space weather research. In this study, we present the ELlipse Evolution model based on HI observations (ELEvoHI) and introduce a deformable front to this model. ELEvoHI relies on heliospheric imagers (HI) observations to obtain the kinematics of a CME. With the newly developed deformable front, the model is able to react to the ambient solar wind conditions during the entire propagation and along the whole front of the CME. To get an estimate of the ambient solar wind conditions, we make use of three different models: Heliospheric Upwind eXtrapolation model (HUX),  Heliospheric Upwind eXtrapolation with time dependence model (HUXt), and EUropean Heliospheric FORecasting Information Asset (EUHFORIA). We test the deformable front on a CME first observed in STEREO-A/HI on February 3, 2010 14:49 UT. For this case study, the deformable front provides better estimates of the arrival time and arrival speed than the original version of ELEvoHI using an elliptical front. The new implementation enables us to study the parameters influencing the propagation of the CME not only for the apex, but for the entire front. The evolution of the CME front, especially at the flanks, is highly dependent on the ambient solar wind model used. An additional advantage of the new implementation is given by the possibility to provide estimates of the CME mass.
\end{abstract}

\section{Introduction} \label{sec:intro}

Coronal mass ejections (CMEs) are large clouds of energetic and magnetized plasma erupting from the solar corona \citep{Hundhausen1994}. They propagate in the solar system and are responsible for the strongest space weather effects. 
Earth directed CMEs can directly impact various systems including space missions, power grids, navigation systems and oil pipelines. \citep[e.g.][]{Gosling1990,Kilpua2012,RichardsonCane2012,Cannon2013}. Therefore, predicting the arrivals of CMEs has become essential. To obtain accurate space weather forecasting it is important to understand the behavior of CMEs in interplanetary space. Furthermore, the properties of CMEs at the time of impact determine the severity of geomagnetic storms \citep{Pulkkinen2007}. These properties are the magnetic field, especially the $B_z$ component, but the size and kinematics of CMEs are also important. It is necessary to understand how CMEs evolve during their propagation in the heliosphere and how they interact with the ambient solar wind to achieve accurate forecasts \citep[e.g.][]{Manchester2017,Kilpua2019}.

Our current real-time CME arrival predictions are not better than $\sim$~10~$\pm$~20~hours \citep{Riley2018}. Today, a large number of CME arrival time and speed forecasting models are available. Table 1 in \cite{Riley2018} lists most of the available models, which exhibit various levels of complexity. For example, the Effective Acceleration Model \citep[EAM;][]{Paouris2017}, uses an empirical relation for the acceleration as a function of the initial speed of the CME. Other models consider physics-based equations and account for drag, i.e.\ drag-based models, between the ambient solar wind and the CME (e.g.\ DBM; \citealt{Vrsnak2013}, DBEM; \citealt{Dumbovic2018}, ANTEATR; \citealt{Kay2020}). Fixed-phi fitting \citep[FPF;][]{Sheeley1999,Rouillard2008}, harmonic mean fitting \citep[HMF;][]{Lugaz2010,Moestl2011}, and self-similar-expansion fitting \citep[SSEF;][]{LugazEtAl2010,Davies2012,MoestlDavies2013} are examples of CME arrival prediction models using wide-angle white light observations from heliospheric imagers (HI) that require techniques assuming certain shapes of the CME front in the ecliptic plane. 
Furthermore, there are prediction models combining both the drag-based approach and HI observations (e.g.\ DBM fitting; \citealt{Zic2015}, Ellipse Evolution model based on HI observations, ELEvoHI; \citealt{Rollett2016, Amerstorfer2018}). Numerical models solve magnetohydrodynamic (MHD) equations, based on synoptic photospheric magnetic-field maps, and simulate the ambient solar wind in the full heliosphere  (e.g., ENLIL; \citealt{Odstrcil2004}, EUHFORIA; \citealt{Pomoell2018}). To provide CME arrival predictions at different locations in the heliosphere, CMEs are injected in the ambient solar wind.

However, none of these models were found to outperform all others \citep{Riley2018}. Some questions arise: What are the main factors that lead to better CME arrival predictions and can we improve forecasts by combining different model approaches?

It has been shown that CMEs may be influenced by different phenomena in the heliosphere, e.g.\ magnetic forces close to the Sun, other CMEs, or by high-speed solar wind streams \citep{Lugaz2012,KayOpher2015,Moestl2015,Shen2011,GUI2011}. The kinematic and morphological characteristics of CMEs can additionally be affected by the ambient solar wind \citep[e.g.][]{Gosling1990,Gopalswamy2000,Manoharan2004, temmer2011,Wang2016,Zhuang2017}. 
CMEs propagating slower than the ambient solar wind speed are likely to experience acceleration while fast CMEs may decelerate \citep{RichardsonCane2010,Manoharan2011}. As a consequence, not only the propagation direction but also the kinematics and shape of CMEs can be altered \citep[e.g.][]{Savani2010,Zuccarello2012,Liu2014,Rollett2014,Ruffenach2015,KayNievesChinchilla2020}.

HI-based prediction models typically assume a certain geometry for the propagation in the heliosphere. In a series of three papers \citep{HowardTappin2009_1, TappinHoward2009_2, HowardTappin2009_3} the authors proposed a model based on the Solar Mass Ejection Imager (SMEI) to constrain the CME frontal shape at large distances from the Sun and to obtain the kinematics of CMEs. The Tappin-Howard (TH) model was further updated to use STEREO data and \cite{HowardTappin2010} showed the applicability for space weather forecasting. \cite{Rollett2014} and \cite{Barnard2017} proposed to include a non-uniform evolution of a CME in order to account for different ambient solar wind conditions. This result is further supported in a statistical study by \cite{Hinterreiter2021}. The authors apply the ELEvoHI method, which assumes an elliptical shape of the CME front and show that predictions for the same CME based on STEREO-A and STEREO-B observations exhibit the largest differences in highly structured ambient wind conditions.

In this study we present the next step in the ELEvoHI model development and account for a time- and spatial dependent drag along the CME front and during the entire propagation of the CME.
With this approach, we aim to shed light upon CME propagation in the interplanetary space by considering different parameters crucial for the arrival time and speed at different locations in the heliosphere.

In Section~\ref{Sec:Data}, we present the selected CME for this case study and list the applied data from different spacecraft. Section~\ref{sec:Methods} deals with ELEvoHI, its set-up and the input data needed as well as the three ambient solar wind models used. In Section~\ref{sec:DeformableFront}, we explain the implementation of the deformable front into ELEvoHI. Section~\ref{sec:Results} lists our results and compares the deformable front to the elliptical front for one event based on the ambient solar wind models. We summarize and discuss our results in Section~\ref{Sec:Summary}.

\section{Data}\label{Sec:Data}

In this case study, we model the arrival time and arrival speed of the CME that hit Earth on February 7, 2010 18:04 UT using ELEvoHI. To run the model we make use of several data products. Most important are images from HI onboard STEREO \citep{Eyles2009}. The HI instrument on each STEREO spacecraft consists of two white-light wide-angle imagers, HI1 and HI2. HI1 has a field-of-view (FOV) extending from 4\textdegree\ -- 24\textdegree\ elongation (angle from Sun center) in the ecliptic and 
HI2 has an angular FOV extending from 18.8\textdegree\ -- 88.8\textdegree\ elongation in the ecliptic. The nominal cadence of the HI1 and HI2 science data is 40 minutes and 120 minutes, respectively. The science image bin size is 70 arc sec for HI1 and 4 arc min for HI2. The studied CME was first observed in STEREO-A/HI on February 3, 2010 14:49 UT. This time corresponds to the unique identifier and time according to the \href{https://www.helcats-fp7.eu}{HELCATS HICAT CME} catalog (version 6). The first observation in STEREO-B occurred six hours later on February 3, 2010 20:49 UT. The HELCATS catalog provides the initial speed of $\sim 350$~km s$^{-1}$ based on self-similar expansion fitting. The CME fronts were tracked by the authors from about 4\textdegree\ to 28\textdegree\ in STEREO-A and from about 6\textdegree\ to 27\textdegree\ in STEREO-B HI observations using ecliptic time-elongation maps \citep[][]{Sheeley1999, Davies2009}. To extract the time-elongation profiles, we use the \href{https://hesperia.gsfc.nasa.gov/ssw/stereo/secchi/idl/jpl/satplot/SATPLOT_User_Guide.pdf}{SATPLOT} tool implemented in IDL\textsuperscript{TM} SolarSoft, which allows any user to measure the elongation at different latitudes. The time-elongation profiles are then converted to time-distance profiles using the ELlipse Conversion \citep[ELCon; a derivation can be found in][]{Rollett2016} procedure. ELCon is similar to other conversion methods (e.g. Fixed-Phi, Harmonic Mean, Self-similar Expansion), but additionally to the propagation direction and longitudinal extent also the shape of the modeled CME front is taken into account. 

\begin{figure}[htbp]
\centering
\includegraphics[width=0.9\linewidth]{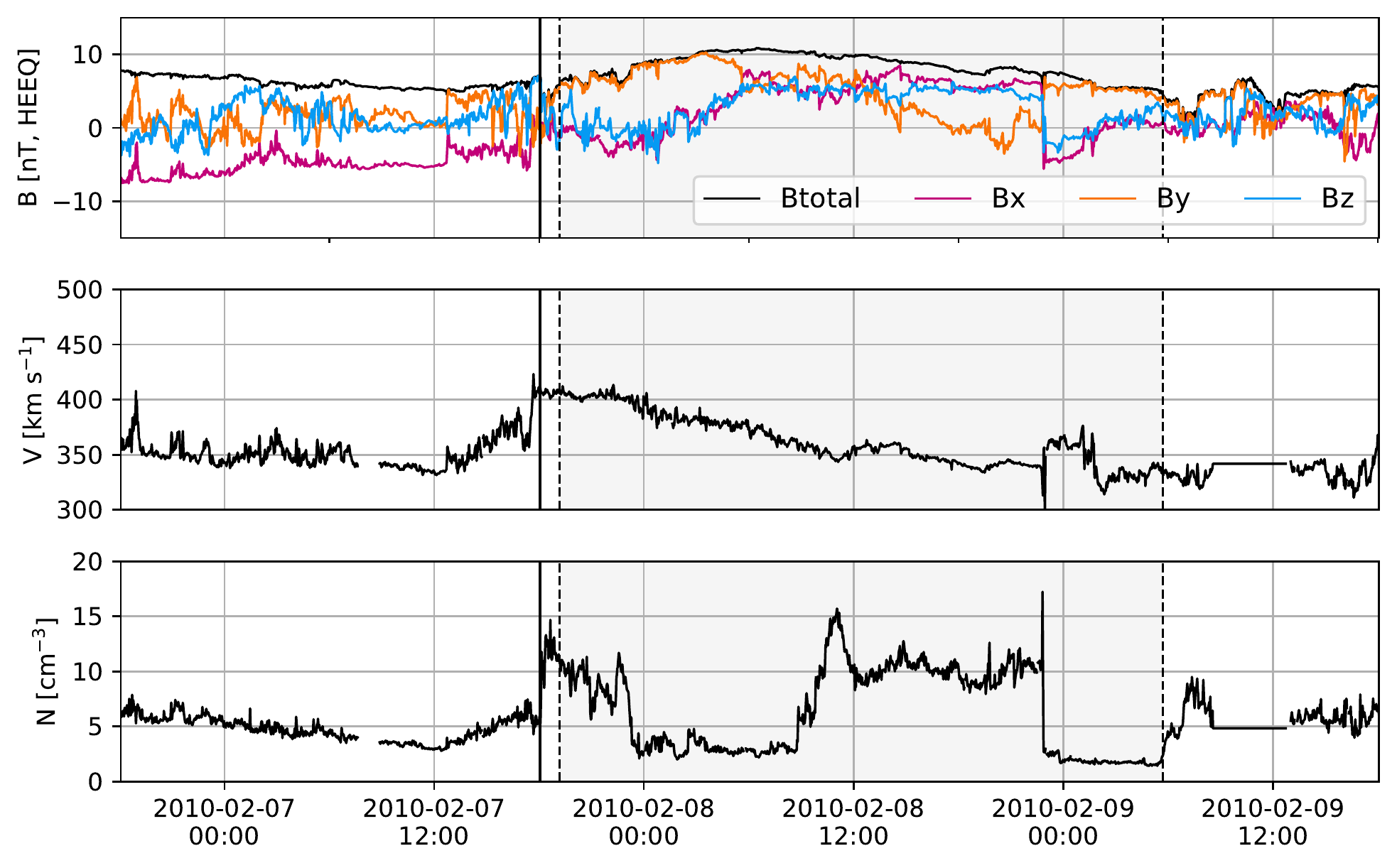}
\caption{In situ signatures of the studied CME. The vertical solid black line indicates the defined arrival time of the CME, which is February 7, 2010 18:04 UT. The vertical dashed black lines define the start and the end time of the magnetic flux rope. The top panel shows the total magnetic field and the individual components. The middle and the lower panel show the solar wind speed and density at Wind spacecraft, respectively.
} \label{fig:ICMEplot}
\end{figure}

Figure~\ref{fig:ICMEplot} shows the in situ solar wind parameters measured by the Wind spacecraft from February 6 -- 9, 2010. Plotted from top to bottom are: the magnetic field components with the total field, the solar wind speed, and solar wind density. The identified interplanetary CME (ICME) in situ arrival time is indicated by the vertical solid black line, while the vertical dashed black line is the start date of the magnetic flux rope. 
The ICME in situ signatures reveal a density enhancement but no shock about 1 hour ahead of a magnetic flux rope (MFR). This density enhancement is used to define the arrival time at Earth, on February 7, 2010 18:04 UT, with an arrival speed of $406\pm2$~km s$^{-1}$. The ICME times and speeds are taken from the HELCATS ICMECAT catalog \citep[version 2.0;][see also the links in the data section]{Moestl2020}, which gives an in situ arrival time of the ICME in question at the Wind spacecraft located in a Lissajous orbit around Lagrange point 1. 

To get the propagation direction and the half width of the CME we use the Ecliptic cut Angles from GCS for ELEvoHI tool \citep[EAGEL,][]{Hinterreiter2021}, which incorporates the Graduated Cylindrical Shell method \citep[GCS,][]{Thernisien2006,Thernisien2009}. Figure~\ref{fig:LatExtent} shows STEREO-A coronagraph images used to perform GCS fitting. STEREO/COR2 have a FOV from 2 -- 15~R$_{\odot}$ with a cadence of the coronagraph science images of about 15 minutes. GCS fitting was performed based on COR2 images from both, STEREO-A and STEREO-B spacecraft (no LASCO data available for this event), on February 3, 2010 15:54 UT. At this time, the CME front was clearly visible and already far out in the coronagraph images. The GCS fitting parameters in Stonyhurst coordintate system are: longitude 355\textdegree, latitude: $-$17\textdegree, tilt angle: $-$1\textdegree, aspect ratio: 0.33, half angle: 30\textdegree. Based on the ecliptic cut, the half width used in this study is 40\textdegree, and the CME propagation direction is set to 68\textdegree\ with respect to STEREO-A, which corresponds to 4\textdegree\ East of Earth. These values serve as initial input to ELEvoHI. The STEREO-A/COR2 images are further used to get an estimate of the latitudinal extent of the CME (see Figure \ref{fig:LatExtent}).

\section{Methods}\label{sec:Methods}
In the following paragraphs, we describe the ELEvoHI ensemble model and the input data needed to obtain an estimate of the arrival time and speed at any location in the heliosphere (Section \ref{sec:ELEvoHI}). An essential input to the model is the ambient solar wind speed in the ecliptic. We therefore employ three different ambient solar wind models, introduced in Section \ref{sec:AmbientSolarWindModels}. The implementation of the deformable front in ELEvoHI not only requires the solar wind bulk speed but also the solar wind mass density, both as a function of radial distance and in the ecliptic plane (Section~\ref{sec:DeformableFront}).
For the CME, we assume the longitudinal and latitudinal expansion to be constant as well as a constant mass during the whole propagation in the heliosphere.

\subsection{Ambient Solar Wind models}\label{sec:AmbientSolarWindModels}
The three ambient solar wind models considered in this study are the Heliospheric Upwind eXtrapolation model \citep[HUX;][]{reiss19,Reiss2020ApJ}, the Heliospheric Upwind eXtrapolation with time dependence model \citep[HUXt;][]{Owens2020SoPh}, and EUropean Heliospheric FORecasting Information Asset \citep[EUHFORIA;][]{Pomoell2018}, which exhibit some differences. HUX and HUXt are based on the solution of the 1D incompressible hyrdrodynamics equations, whereas EUHFORIA is based on the solution of the full 3D MHD equations. Additionally, HUX and EUHFORIA provide a static solution of the ambient solar wind for a full Carrington rotation, HUXt provides a map of the ambient solar wind speed for each time step. Important for the deformable front is an estimate not only for the ambient solar wind speed but also for the ambient solar wind density. Contrary to the other two models, EUHFORIA self-consistently models the plasma dynamics and thus also provides the ambient solar wind density, $n$. For HUX and HUXt, we rely on an empirical relation proposed by \cite{EyniSteinitz1980}: 

\begin{equation}
\label{Eq:ES1980}
    n(r, w)=1.3\times10^6  r^{-2.0}  w^{-2.0},
\end{equation}

where $r$ [AU] is the radial distance and $w$ [km~s$^{-1}$] the solar wind speed. Hence, $n$, [protons~cm$^{-3}$] is not only dependent on the radial distance to the Sun but also on the ambient solar wind speed, leading to a structured ambient solar wind density.

\subsubsection{HUX}\label{HuxModel}

To model the physical conditions in the evolving ambient solar wind flow, we use the numerical framework discussed in~\cite{reiss19, Reiss2020ApJ}. We specifically use magnetic maps of the photospheric magnetic field from the Global Oscillation Network Group (GONG) provided by the National Solar Observatory (NSO) as input to magnetic models of the corona. Using the Potential Field Source Surface model~\citep[PFSS;][]{altschuler69, schatten69} and the Schatten current sheet model~\citep[SCS;][]{schatten71} we compute the global coronal magnetic field topology. While the PFSS model attempts to find the potential magnetic field solution in the corona with an outer boundary condition that the field is radial at the source surface at 2.5~R$_{\odot}$, the SCS model in the region between 2.5 and 5~R$_{\odot}$ accounts for the latitudinal invariance of the radial magnetic field as observed by Ulysses \citep[][]{wang95}. From the global magnetic field topology, we calculate the solar wind conditions near the Sun using the established Wang-Sheeley-Arge (WSA) relation~\cite{wang95, arge03, riley11b} as described in \cite{reiss19}. 
To evolve the solar wind solutions from near the Sun to Earth, we use the Heliospheric Upwind eXtrapolation model (HUX)~\cite{riley11b}. The HUX model simplifies the fluid momentum equation as much as possible, by neglecting the pressure gradient and the gravitation term in the fluid momentum equations as proposed by \cite{riley11b}. The model solutions match the dynamical evolution explored by global heliospheric MHD codes fairly well while having low processor requirements. 

HUX provides a static solution of the ambient solar wind for a full Carrington rotation. The data spans from 5 to 430~R$_{\odot}$ with a radial resolution of 1~R$_{\odot}$ while the longitudinal resolution is 2\textdegree.

\subsubsection{HUXt}\label{HUXt}
HUXt is a solar wind numerical model that treats the solar wind as a 1D incompressible, time-dependent hydrodynamic flow \citep{Owens2020SoPh}. This reduced physics approach enables very efficient computational solutions, which are approximately 10$^3$ times faster than comparable 3D MHD solar wind solutions. Nonetheless, HUXt can closely emulate the solar wind speed output of full 3D MHD solar wind models \citep{Owens2020SoPh}. Consequently, HUXt can be a useful surrogate in situations where full 3D MHD solar wind simulations are too computationally expensive - for example, large ensemble simulations \citep{Barnard2020}. The only boundary condition of HUXt is the solar wind speed on the inner boundary, which is typically derived from the output of coronal models.

For this study we use the HUXt model with the inner boundary conditions from WSA, provided by the 
\href{https://ccmc.gsfc.nasa.gov/results/viewrun.php?domain=SH&runnumber=Tanja_Amerstorfer_041521_SH_1}{CCMC}. HUXt data starts at 21.5~R$_{\odot}$, corresponding the outer boundary from the WSA, and reaches up to 300.5 R$_{\odot}$ with a resolution of 1~R$_{\odot}$. The longitudinal resolution is 0.7\textdegree\ while the temporal resolution is given by 3.865 minutes. 

\subsubsection{EUHFORIA}\label{EUHFORIA}
As noted in the previous sections, EUHFORIA models the dynamical evolution of the solar wind in the inner heliosphere by numerically solving the equations of single-fluid magnetohydrodynamics (including gravity) in a three-dimensional volume starting at a heliocentric distance of 0.1~AU. On the sphere defining the inner radial boundary, the MHD quantities representing the solar wind at that heliocentric distance need to be specified. This is most often done by employing empirical relations that are based on magnetic field models of the low and extended corona using the PFSS and SCS models, respectively.
For this study, as input to the coronal model, a synoptic magnetogram constructed 
from SOHO/MDI observations for Carrington rotation 2093 as provided by the Joint Science Operations Center (JSOC) was used.

To arrive at a solution describing the heliospheric plasma conditions at a given time, EUHFORIA solves the MHD equations in the HEEQ coordinate frame until a steady-state solution in the co-rotating frame is achieved. Thus, after this time, if the boundary conditions do not evolve in this frame, the solution remains unchanged. Employing this assumption in this study, the solar wind conditions  
like for HUX, are provided as a steady-state solution for a full Carrington rotation. The model output spans from 20.56 to 324.43~R$_{\odot}$ with a resolution of 0.94~R$_{\odot}$ while the longitudinal and latitudinal resolution is 1\textdegree. EUHFORIA not only provides the ambient solar wind speed but all MHD quantities and therefore self-consistently provides the ambient solar wind density. Note that for this study, from the model output a two-dimensional slice of data representing the ecliptic plane is henceforth used in all the analysis. 

\subsection{ELEvoHI ensemble modeling}\label{sec:ELEvoHI}
ELEvoHI uses HI time-elongation profiles of CME fronts and assumes an elliptical shape for those fronts to derive their interplanetary kinematics. The model converts the resulting time-elongation profiles to time-distance profiles, assuming an elliptic frontal shape using the ELEvoHI built-in procedure ELCon. Furthermore, ELEvoHI accounts for the effect of the drag force exerted by the ambient solar wind. The interaction of the CME with the solar wind, that can effectively be described by introducing a drag term in the equation of motion, is an essential factor influencing the dynamic evolution of CMEs in the heliosphere. ELEvoHI incorporates a drag-based equation of motion \cite[DBM;][]{Vrsnak2013} to fit the time-distance tracks. 
Within these profiles, the user has to manually define the start- and end point for the DBM fit. For this event they are set to around $30$~R$_{\odot}$ and $65$~R$_{\odot}$, respectively. In order to account for the de-/acceleration of the CME due to drag, an estimate of the ambient solar wind speed is needed. 

In a previous study by \cite{Amerstorfer2021}, the authors applied different approaches to get an estimate of the ambient solar wind speed used as input to ELEvoHI. They tested 1) the ambient solar wind speed from the HUX model, 2) a range of possible solar wind speeds (225 -- 625~km~s$^{-1}$), and 3) solar wind speed measured at L1 during the evolution of the CME, and found the best results based on the HUX ambient solar wind conditions.

In this study we make use of three different ambient solar wind models: HUX, HUXt, and EUHFORIA. The ambient solar wind speeds in the ecliptic plane for each model can be seen in Figure~\ref{fig:DefFront}, with snapshots of the ELEvoHI modeled CME fronts. The estimate of the ambient solar wind speed used for DBM fitting is obtained identically for each model.
We only consider the region of the full ambient solar wind speed data according to the start- and end-point selected by the user, the CME propagation direction, and the half width for each ensemble member. This corresponds to the radial extent used for DBM fitting \cite[see Section 3.3 in][]{Hinterreiter2021}. From that region we take the median of the solar wind speed and define the uncertainties to be $\pm$100~km~s$^{-1}$, based on a study by \cite{Reiss2020ApJ}, where the authors considered nine years (mid 2006 to mid 2015) and report a mean absolute error of the HUX solar wind speed prediction with respect to the in situ speed of 91~km~s$^{-1}$ \cite[see Section 3.3 in][for more details]{Hinterreiter2021}. For consistency, we also apply the same uncertainties for the obtained median solar wind speed for the HUXt and the EUHFORIA ambient solar wind models.
We then split the ambient solar wind speed with its uncertainty into steps of 25~km~s$^{-1}$, leading to nine different input speeds to ELEvoHI. For each of the nine input speeds DBM fitting is performed. ELEvoHI then selects the combination of drag parameter and ambient solar wind speed that best fits the time-distance profile for each ensemble member \citep[for a detailed description see][]{Rollett2016}. 

The selected drag parameter, $\gamma$, and solar wind speed, $w$, from DBM fitting are assumed to be valid for the entire propagation of the apex, which is defined by Equation \ref{Eq:v} and Equation \ref{Eq:r} \citep{Vrsnak2013}:
\begin{equation}
\label{Eq:v}
    v(t)=\frac{v_0-w}{1\pm\gamma(v_0-w)t}+w
\end{equation}
\begin{equation}
\label{Eq:r}
    r(t)=\pm \ln[1\pm\gamma(v_0-w)t]+wt+r_0, 
\end{equation}
with $v_0$ as the initial CME speed while $t$ defines the time of the CME propagation. An important factor in these equations is the sign of $\gamma$. It is defined so that the CME accelerates when the sign is negative while the CME front decelerates when the sign of $\gamma$ is positive. 

In order to get the shape and the propagation direction of the CME we make use of the EAGEL tool \citep{Hinterreiter2021}. It provides the propagation direction with respect to the observer ($\phi$ = 68\textdegree, with respect to STA) and half width ($\lambda$ = 40\textdegree). The inverse ellipse aspect ratio, $f$, defines the shape of the assumed CME front in the ecliptic plane, where $f=1$ represents a circular front, while $f<1$ corresponds to an elliptical CME front (with the semi-major axis perpendicular to the propagation direction).

ELEvoHI is operated in ensemble mode by varying $\phi$, $\lambda$, and $f$ \cite[for a detailed description see][]{Amerstorfer2018}. The parameters $\phi$ and $\lambda$ vary over a range of $\pm10$\textdegree\ with a step size of $2$\textdegree\ and $5$\textdegree, respectively.
The range $\pm10$\textdegree\ is based on a study by \cite{Mierla2010}, in which the authors report an uncertainty in the parameters when different users manually perform GCS reconstruction. For $f$ we set a fixed range from $0.7-1.0$ ($0.1$ step size). Thus we get a total of 220 ensemble members for one event (i.e.\ 11 values of $\phi$, 5 values of $\lambda$ and 4 values of $f$). 
When running ELEvoHI in ensemble mode, we get a frequency distribution from which we can calculate the median, mean and standard deviation of the modeled CME arrival time and speed.
In addition, we can give a probability for whether a CME is likely to hit Earth or not. When all of the 220 ensemble members model an arrival at Earth, we assume the likelihood of an Earth hit to be 100\%.

\subsection{Implementation of the deformable CME front}\label{sec:DeformableFront}
In the original version of ELEvoHI, i.e.\ for the elliptical front, the apex of the CME propagates the whole way through the heliosphere according to the ambient solar wind speed and drag parameter obtained from DBM fitting. 

For the deformable front, however, $\gamma$ and $w$ from the DBM fit are not considered for the entire propagation of the CME front, but only up to about 65~R$_{\odot}$ (corresponding to the endcut of the DBM fit defined by the user). At this distance we start a transition from the rigid elliptical front to a deformable front. We define the front to consist of 101 points, leading to a longitudinal resolution of about 1\textdegree\ when assuming a half width of 50\textdegree. With decreasing $\lambda$ the longitudinal resolution increases. Each point of the front can propagate individually according to the different ambient solar wind conditions. We therefore need to know the parameters in Equation \ref{Eq:v} and \ref{Eq:r} ($v_0$, $w$, $\gamma$) at each time and location in the heliosphere. The CME frontal speed for each point, $v_0$, is obtained from the previous time step, while the solar wind speed, $w$, for each time and location is taken from the ambient solar wind models. To derive the drag parameter, $\gamma$, for each time and location we have to make further assumptions. That is, the longitudinal and latitudinal expansion as well as the mass, $M$, of the CME is constant during the entire propagation.

In order to obtain an estimate of $M$, we use a similar approach as \cite{Amerstorfer2018} and rearrange Equation~\ref{Eq:gamma} \citep{Cargill2004}: 
\begin{equation}
\label{Eq:gamma}
    \gamma(r)=c_d\frac{A(r)n(r,w)}{M},
\end{equation}

where $\gamma$ is the drag parameter, $c_d$ is a dimensionless drag coefficient and is set to 1 in this study. $A$ is the cross-sectional area of the CME, $n$ is the ambient solar wind density. We get $\gamma$ and $w$ from DBM fitting, i.e.\ the drag parameter and the ambient solar wind at the transition from rigid to deformable front. Also the radial distance of the front at this time is known, so $n(r,w)$ can be derived from Equation~\ref{Eq:ES1980} and $A(r)$ can be calculated (see below). Note that $n$ is provided by EUHFORIA and can therefore directly be used within ELEvoHI. An estimate of the CME mass can now be given based on DBM fitting. Furthermore, $\gamma$ can be expressed by the radial distance and the solar wind density at any location in the heliosphere, by assuming a constant mass.

\begin{figure}[htbp]
\centering
\includegraphics[width=1.0\linewidth]{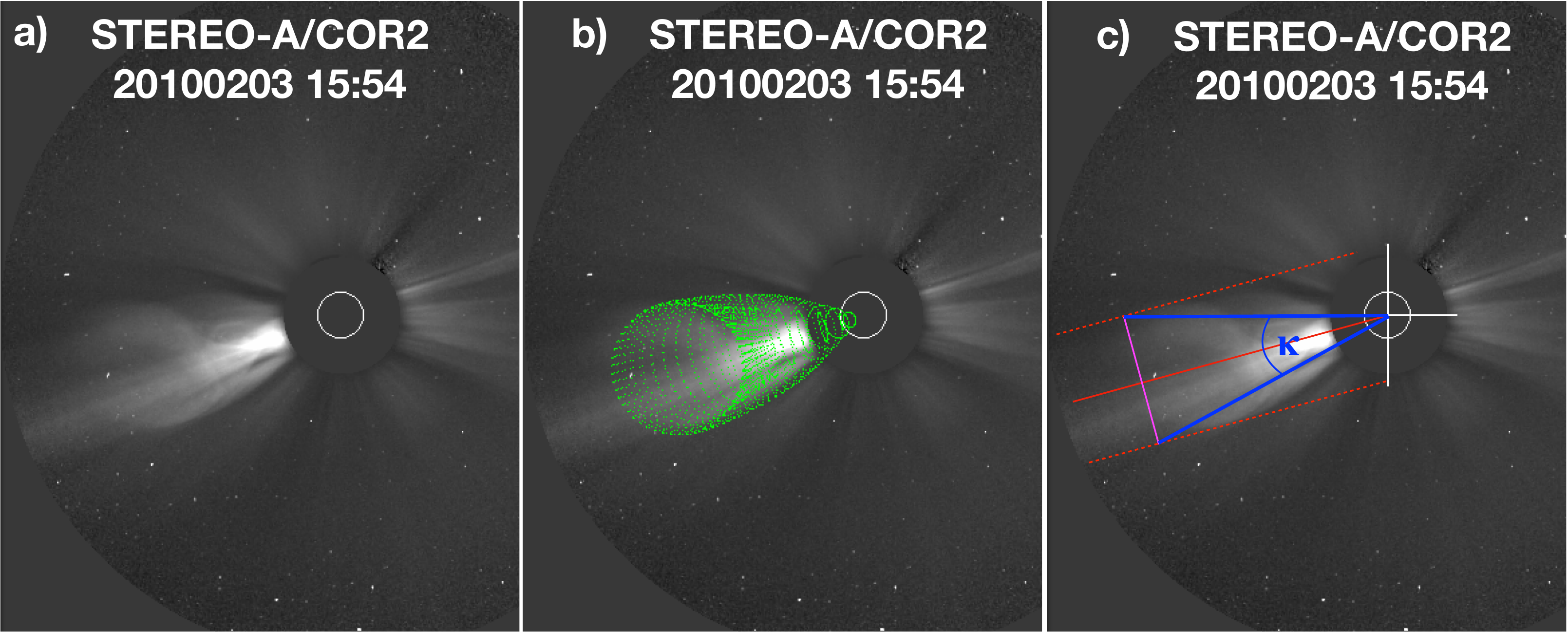}
\caption{STEREO-A coronagraph images for the CME on February 3, 2010. a) COR2 image at 15:54 UT. b) Same as a) with the GCS wireframe overplotted. c) COR2 image with the definition of the latitudinal extent of the CME. The red dashed lines represent the maximum extent (north and south) of the CME as viewed from the propagation direction in the latitude (solid red line). The solid magenta line defines the CME front. The angle ($\kappa$) between the solid blue lines represents the latitudinal extent of the CME.} \label{fig:LatExtent}
\end{figure}

To get an estimate of the cross-sectional area, $A$, at different time steps of the model, we assume a constant expansion in longitude and latitude. The longitudinal extent of the CME is obtained by EAGEL and is defined by $\lambda$. For the latitudinal extent, we make use of STEREO coronagraph images (see Figure \ref{fig:LatExtent}). We first define the main latitudinal propagation direction (red solid line in Figure \ref{fig:LatExtent}c). Next, two parallel lines are added at the maximum northern and southern extent of the CME (dashed red lines in Figure \ref{fig:LatExtent}c). The magenta line is orthogonal to the red lines and indicates the CME front. The intercept of the magenta line with the dashed red lines represents the maximum latitudinal extent of the CME. 
The blue solid lines connect the two intercepts with the solar center and therefore provide an angle ($\kappa$) for the latitudinal extent of the CME ($\kappa$~=~28\textdegree\ for this event). As mentioned above, $\kappa$ is assumed to be constant during the propagation.
In good approximation, the cross-sectional area can be considered as an ellipse ($A=a b \pi$). The semi major axis, $a$, is defined by $\lambda$ and can be calculated for each radial distance from the Sun. The same applies for the semi minor axis, $b$, which is dependent on $\kappa / 2$ and the radial distance. As a consequence, $A$ can be expressed with regard to the radial distance of the CME front to the Sun, i.e.\ $A = A(r)$.

With the assumptions mentioned previously, all the parameters in Equation \ref{Eq:v} and \ref{Eq:r} at any time and location in the heliosphere can be estimated. So, at around 65~R$_{\odot}$ we perform a transition from the rigid elliptical CME front to the deformable front that is able to react to the different solar wind conditions. We set this distance in agreement with \cite{Owens2017Nat}, who found that at about 0.3~AU the majority of CMEs can no longer be considered as coherent structures.
We set the temporal resolution for the deformable front to 15 minutes. Only for HUXt the temporal resolution is set to be 15.46 minutes, which corresponds to 4 times the temporal resolution of the model output. 

Note that the results for the rigid elliptical front are still generated, allowing us to compare the modeled arrivals for the different implementations of the ELEvoHI.

\section{Results} \label{sec:Results}

\begin{figure}[htbp]
\centering
\includegraphics[width=0.64\linewidth]{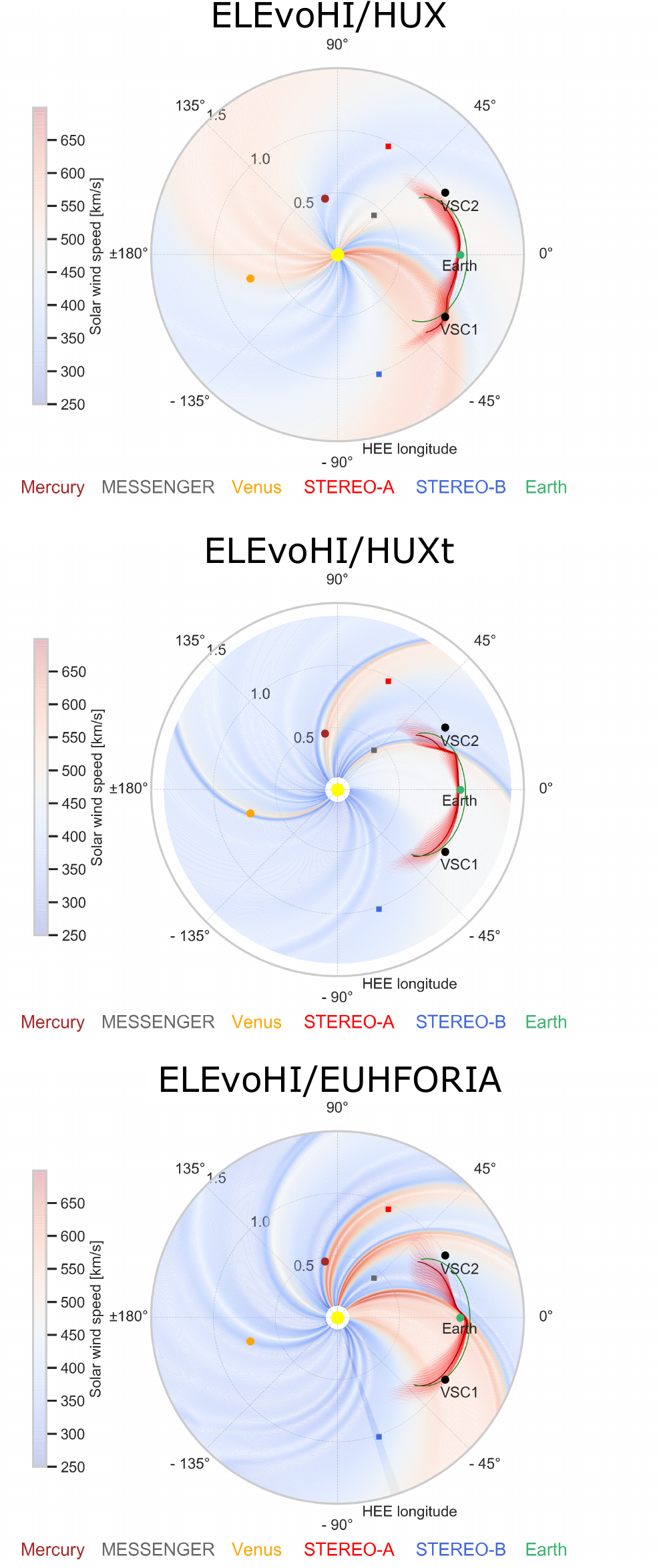}
\caption{Snapshots of the ELEvoHI model results. From top to bottom the CME fronts based on HUX, HUXt, and EUHFORIA are shown. The green solid line represents the elliptical CME front (for one individual ensemble member) and the red lines represent the deformed fronts. The dark red line corresponds to the same individual run as for the elliptical CME front (green line). Plotted in black are the positions of the virtual spacecraft (VSC1 and VSC2), which are located $\pm$30\textdegree\ East and West of Earth. The positions of additional planets and spacecraft are indicated by the colored circles and squares, respectively.} \label{fig:DefFront}
\end{figure}

Figure \ref{fig:DefFront} shows one ensemble member of the elliptical front (green) and all the ensemble members of the deformed front (red) for the three different ambient solar wind models used as input. The dark red deformed front corresponds to the single ensemble member shown in green for the elliptical front. The ELEvoHI input parameters for this ensemble member are: $\phi$ = 68\textdegree\ with respect to STEREO-A (corresponding to 4\textdegree\ with respect to Earth), $\lambda$ = 40\textdegree\, and $f$ = 0.7.
In Table \ref{tab:EnsembleResults} we list the modeled arrival times for the elliptical and the deformed front for the three ambient solar wind models. Note that all of the individual ensemble members estimate an arrival at Earth giving a 100\% chance of an Earth hit.
Table \ref{tab:EnsembleResults} further lists the modeled arrival times at two different predefined positions in the heliosphere, called virtual spacecraft (VSC). VSC1 and VSC2 are located $\pm$30\textdegree\ East and West of Earth, respectively. We include these two additional locations in order to assess the CME propagation at the flanks. Furthermore, introducing VSC1 and VSC2 allows us to point out the differences based on the three ambient solar wind models at other longitudes.
In contrast to the 100\% chance of an arrival at Earth, not all ensemble members are estimated to arrive at VSC1 and VSC2. The reason can be found in the changing propagation direction and half width for each of the ensemble members. 

\begin{table}\label{tab:Arrivaltimes}
\centering
\caption{Modeled arrival times for different ambient solar wind models and locations for the elliptical and the deformed CME front. Given are the median arrival times with the standard deviation as uncertainty. $\Delta_{\mathrm{insitu}}$ lists the difference to the in situ arrival time for both the elliptical and deformed front. $\Delta_{\mathrm{shape}}$ gives the difference between the two frontal shapes, where a positive value represents a later arrival of the deformed front. The in situ arrival time is defined to be February 7, 2010 18:04 UT.}

\begin{tabular}{cccccc}
\hline 
Location & AT$_{\mathrm{ellipse}}$ & $\Delta_{\mathrm{insitu}}$ & AT$_{\mathrm{deformed}}$ & $\Delta_{\mathrm{insitu}}$ & $\Delta_{\mathrm{shape}}$ \\
& [UT $\pm$ h] & [h] & [UT $\pm$ h] & [h] & [h]\\
\hline 
\hline 
\multicolumn{4}{|l|}{\textbf{ELEvoHI/HUX}}\tabularnewline
\hline 
Earth & 2010-02-07 10:54 $\pm$ 0.7 & -7.2 & 2010-02-07 16:21 $\pm$ 0.6 & -1.7 & 5.5 \tabularnewline
VSC1 & 2010-02-07 22:44 $\pm$ 10.2 & --- & 2010-02-07 17:51 $\pm$  3.0 & --- & -4.9 \tabularnewline
VSC2 & 2010-02-08 05:24 $\pm$ 9.7 & --- & 2010-02-08 04:06 $\pm$ 3.7 & --- & -1.3 \tabularnewline
\hline
\hline
\multicolumn{4}{|l|}{\textbf{ELEvoHI/HUXt}}\tabularnewline
\hline 
Earth & 2010-02-07 12:04 $\pm$ 0.6 & -6.0 & 2010-02-07 16:26 $\pm$ 0.5 & -1.6 & 4.4 \tabularnewline
VSC1 & 2010-02-08 00:04 $\pm$ 10.2 & --- & 2010-02-08 02:14 $\pm$ 5.2 & --- & 2.1 \tabularnewline
VSC2 & 2010-02-08 06:44 $\pm$ 10.2 & --- & 2010-02-08 14:21 $\pm$ 6.0 & --- & 7.6 \tabularnewline
\hline
\hline
\multicolumn{4}{|l|}{\textbf{ELEvoHI/EUHFORIA}}\tabularnewline
\hline 
Earth & 2010-02-07 09:34 $\pm$ 1.1 & -8.5 & 2010-02-07 11:51 $\pm$ 0.6 & -6.2 & 2.3 \tabularnewline
VSC1 & 2010-02-07 20:39 $\pm$ 10.2 & --- & 2010-02-07 22:29 $\pm$ 5.2 & --- & 1.8 \tabularnewline
VSC2 & 2010-02-08 03:44 $\pm$ 9.2 & --- & 2010-02-08 13:06 $\pm$ 9.0 & --- & 9.4 \tabularnewline
\hline
\end{tabular}\label{tab:EnsembleResults}

\end{table}

\subsection{Model results for the elliptical front}
From Table \ref{tab:EnsembleResults} it can be seen that the elliptical fronts of all of the solar wind models estimate the Earth arrival too early (in situ arrival time is defined to be February 7, 2010 18:04 UT). The modeled arrival times are February 7, 2010 10:54 UT $\pm$ 0.7 hours, February 7, 2010 12:04 UT $\pm$ 0.6 hours, and February 7, 2010 09:34 UT $\pm$ 1.1 hours for ELEvoHI/HUX, ELEvoHI/HUXt, and ELEvoHI/EUHFORIA, respectively. The largest difference within the ambient solar wind models is found for ELEvoHI/HUXt and ELEvoHI/EUHFORIA with 2.5~hours. This leads to more than 8.5~hours difference for the calculated arrival time based on ELEvoHI/EUHFORIA with respect to the actual in situ arrival time. Also the modeled arrival times for the virtual spacecraft, differ up to about 3.5~hours for VSC1 and 3~hours for VSC2.

To find the reasons for the differences, we check the median ambient solar wind speed in the range corresponding to the start- and endcut of the DBM fit of each model. From ELEvoHI/HUX we obtain $455$~km~s$^{-1}$, from ELEvoHI/HUXt it is $421$~km~s$^{-1}$. For ELEvoHI/EUHFORIA the median ambient solar wind speed is $561$~km~s$^{-1}$ (more than $100$~km~s$^{-1}$ faster than for the other two models). The in situ solar wind speed is roughly $500$~km~s$^{-1}$ about 3.5 days prior to the actual arrival and gradually decreases to about $350$~km~s$^{-1}$ (see Figure \ref{fig:SpeedProfile}). When checking the speed from the best DBM fit, we find for ELEvoHI/HUX:~$555$~km~s$^{-1}$, for ELEvoHI/HUXt:~$521$~km~s$^{-1}$, and for ELEvoHI/EUHFORIA:~$661$~km~s$^{-1}$, indicating that ELEvoHI selects the fastest ambient solar wind available. The drag parameters, $\gamma$, are $2.73\times10^{-8}$~km$^{-1}$  for ELEvoHI/HUX, $4.20\times10^{-8}$~km$^{-1}$ for ELEvoHI/HUXt, and $1.07\times10^{-8}$~km$^{-1}$ for ELEvoHI/EUHFORIA. The $\gamma$ obtained for all the models seems to be roughly in the same range of other studies \citep[see, e.g.][]{Vrsnak2013,Dumbovic2018, Rollett2016}. Even with the largest $\gamma$, in this case the highest acceleration, the HUXt based model provides the latest arrival at Earth.

\subsection{Model results for the deformed front}
Next, we compare the modeled arrival times for the deformed front based on the three different ambient solar wind models. Here we find an almost identical modeled arrival time for ELEvoHI/HUX and ELEvoHI/HUXt on February 7, 2010 16:21 UT and 16:26 UT, respectively (see Table \ref{tab:EnsembleResults}). They are about two hours too early with respect to the actual in situ arrival time, while ELEvoHI/EUHFORIA models the arrival time more than 6 hours too early. The calculated arrival times at VSC1 exhibit quite large differences of more than 8.5~hours for ELEvoHI/HUX and ELEvoHI/HUXt. At VSC2 location, the calculated arrival times show even larger differences of more than $\sim$~10~hours.

\begin{figure}[htbp]
\centering
\includegraphics[width=1.0\linewidth]{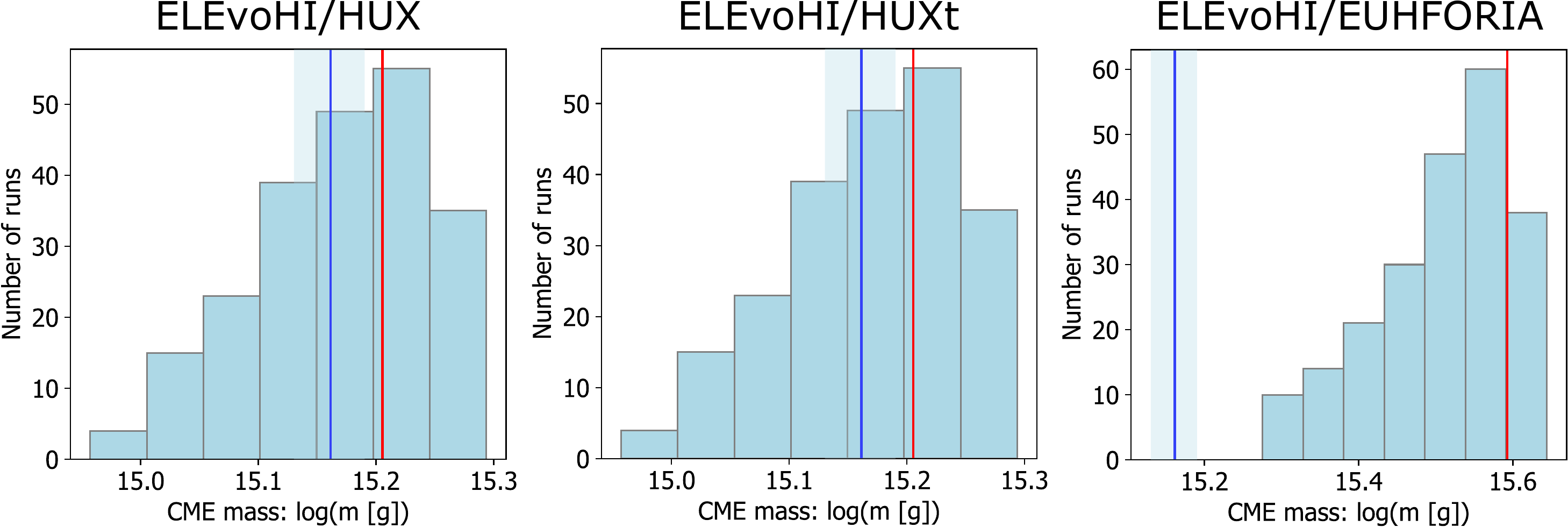}
\caption{Calculated masses for each individual ensemble member and the three ambient solar wind models. The red vertical line represents the mass obtained for the individual ensemble run plotted in dark red in Figure \ref{fig:DefFront}. The blue vertical line indicates the CME mass with its uncertainty obtained from coronagraph images.} \label{fig:MassDistirbution}
\end{figure}

To find the reason for the arrival time variations based on the ambient solar wind models, we check the input parameters to the deformable front right at the transition from the elliptical to the deformed front. The CME speed at the transition is similar based on all the three ambient solar wind models and reaches $404$~km~s$^{-1}$, while a calculated cross-sectional area, $A$, of $6.93\times10^{14}$~km$^{2}$ is obtained. $\gamma$ and $n$ are based on the DBM fit and therefore lead to different values for each ambient solar wind model. When expressing $M$ from Equation \ref{Eq:gamma} we get $1.17\times10^{15}$~g for ELEvoHI/HUX, $1.61\times10^{15}$~g for ELEvoHI/HUXt, and $3.92\times10^{15}$~g for ELEvoHI/EUHFORIA, which is more than two times larger than for the other two models. However, these values are in good agreement with the CME mass estimated based on coronagraph images of $1.45 \pm 0.15\times10^{15}$~g. In coronagraph images, the CME mass is defined via the excess brightness in the white-light image. Assuming a composition of 90\% hydrogen and 10\% helium, the brightness is converted into electron mass (see \citealt{Billings1966}). A detailed description of how the CME mass is estimated can be found in \cite{Colaninno2009} and \cite{Bein2013}, while \cite{DeKoning2017} provides a discussion regarding the uncertainties. In Figure \ref{fig:MassDistirbution} the calculated mass based on the three different ambient solar wind models are shown. The red vertical line indicates the input parameters for the individual run shown in dark red in Figure \ref{fig:DefFront}. For all the input parameters from the ensemble mode to the deformable front see the \href{https://doi.org/10.6084/m9.figshare.14923032.v1}{supplementary material} . 

\subsection{Deformation measure}
In Figure \ref{fig:DefFront} the green solid line represents the ELEvoHI elliptical CME front, while the dark red solid line is the deformed front for one ensemble member. We further aim to find a measure to determine the deformation of the CME front with regard to the elliptical front. To do so, we calculate the mean of the absolute difference in radial coordinate ($\Delta F$) of each point from the elliptical and the deformed CME front at the arrival time at Earth. This gives a first indication on the difference between the elliptical and the deformed front. However, this value is not just dependent on the deformation, but also changes when the deformed front propagates faster or slower than the elliptical front. Hence, we provide an additional parameter, $\sigma F$, which is defined to be the standard deviation of the absolute differences for each point on the CME front. A larger value of $\sigma F$ represents a more deformed CME front. For the single ensemble member (dark red and green lines shown in Figure \ref{fig:DefFront}) of ELEvoHI/HUX, we obtain $\Delta F = 12.1$~R$_{\odot}$ and $\sigma F = 7.3$~R$_{\odot}$. The parameters for ELEvoHI/HUXt are $\Delta F = 9.2$~R$_{\odot}$ and $\sigma F = 4.2$~R$_{\odot}$ and for ELEvoHI/EUHFORIA we obtain $\Delta F = 11.5$~R$_{\odot}$ and $\sigma F = 6.8$~R$_{\odot}$. Based on the $\sigma F$ values for the different ambient solar wind models, the ELEvoHI/HUX results show the largest deformation, followed by the ELEvoHI/EUHFORIA and ELEvoHI/HUXt.
To get an impression for these values, we also calculate these measures only for the elliptical front on February 7, 2010 13:00 UT and 5 hours later (February 7, 2010 18:00 UT) for ELEvoHI/HUX. We find $\Delta F = 11.0$~R$_{\odot}$ and $\sigma F = 0.8$~R$_{\odot}$, indicating that the CME front shows almost no deformation but the absolute difference between the CME points is comparable to the deformed front.

\subsection{Behavior of the propagation parameters}
\begin{figure}[htbp]
\centering
\includegraphics[width=0.9\linewidth]{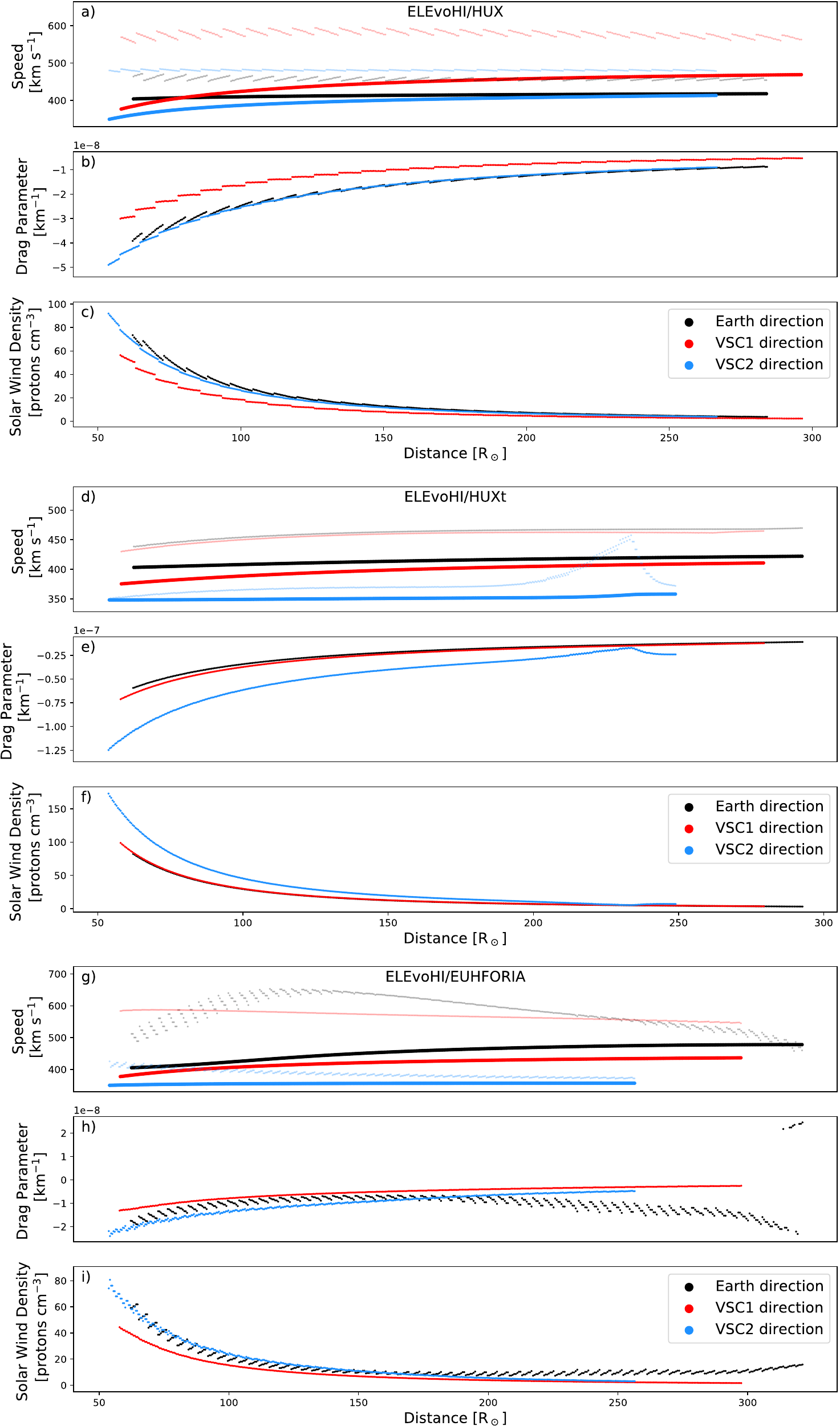}
\caption{Extracted parameters over distance in the heliosphere for the three different ambient solar wind models. The positions are indicated by the different colors, where black represents Earth direction, red represents VSC1, and blue VSC2. ELEvoHI/HUX: panel a), b), c); ELEvoHI/HUXt: panel d), e), f); ELEvoHI/EUHFORIA: panel f), h), i). In panel a), d), and g) the ambient solar wind speed (faint colors) and the speed of the CME front (bold colors) are shown. Panels b), e), h) show the drag parameter and panels c), f), i) the ambient solar wind density.} \label{fig:Behavior}
\end{figure}

Another interesting point is how the individual parameters develop during the propagation of the CME front in the heliosphere. We therefore consider the ambient solar wind speed, the CME frontal speed, the drag parameter, and the ambient solar wind density. In Figure \ref{fig:Behavior} these parameters are plotted for ELEvoHI/HUX, ELEvoHI/HUXt, and ELEvoHI/EUHFORIA, respectively. The plots further show the four parameters for three different propagation directions along predefined longitudes: Earth, VSC1, and VSC2. Earth direction (black) is the longitude corresponding to Earth location. VSC1 (red) and VSC2 (blue) are virtual spacecraft located 30\textdegree\ East and West of Earth, respectively. For the ELEvoHI/HUX Earth direction the ambient solar wind is in the range of $450$~km~s$^{-1}$. The same applies for the ELEvoHI/HUXt Earth direction, while here the ambient solar wind starts slightly below $450$~km~s$^{-1}$. The ambient solar wind speed for ELEvoHI/EUHFORIA shows the largest variation starting from roughly $500$~km~s$^{-1}$, rising to about $650$~km~s$^{-1}$ and coming back to about $500$~km~s$^{-1}$. 

A striking feature in Figure \ref{fig:Behavior} is that the ambient solar wind speed shows 'jumps' for ELEvoHI/HUX and ELEvoHI/EUHFORIA nearly throughout the entire propagation and for almost every longitude plotted. The reason can be found in the static solution of the ambient solar wind speed provided by these models and the temporal resolution of ELEvoHI. In order to select the corresponding ambient solar wind speed at a given time and location in the heliosphere, we purely rotate the solar wind model output according to the correct time. The small 'jumps' in the plot arise from changing from one grid cell to the other in the radial direction, while the large 'jumps' are due to the change from one longitude to the next. The 'jumps' in $\gamma$ and $n$ are due to the 'jumps' in the solar wind speed since these parameters are derived from the solar wind speed. Even though the ELEvoHI/HUXt ambient solar wind model is time dependent (with a resolution of 3.865 minutes) the speeds also exhibit small 'jumps'. They occur, however, only in regions where the ambient solar wind changes significantly during a short period of time (see VSC2 in the HUXt panel in Figure \ref{fig:Behavior}).

For all of the ambient solar wind models the CME frontal speeds, at the three predefined longitudes, do not reach the ambient solar wind speed leading to a continuous acceleration of the front up to L1 distance (roughly 214~R$_{\odot}$). $\gamma$ is quite small for all the models and directions already in the beginning, with the exception of VSC2 direction for ELEvoHI/HUXt. Furthermore, $\gamma$ decreases due to the decreasing ambient solar wind density, $n$, when the front is farther out in the heliosphere. Therefore, it is less likely that the CME catches up with the ambient solar wind farther out in the heliosphere. For ELEvoHI/EUHFORIA however, it can be seen that at about 320~R$_{\odot}$ the CME speed is higher than the ambient solar wind speed. This directly leads to change in sign of $\gamma$ and corresponds to a deceleration of the CME front within Earth direction.

The modeled arrival time for the deformed front shows the largest discrepancy to the actual in situ arrival time for the ELEvoHI/EUHFORIA combination. We believe that this mainly arises from the high ambient solar wind speed. While the Earth-directed part for ELEvoHI/HUX and ELEvoHI/HUXt only slightly accelerates, the modeled speed from ELEvoHI/EUHFORIA increases from about $400$~km~s$^{-1}$ up to more than $475$~km~s$^{-1}$ at the end of the simulation, resulting in an even earlier arrival than for ELEvoHI/HUX and ELEvoHI/HUXt.

\subsection{Modeled CME arrival speed}
\begin{figure}[htbp]
\centering
\includegraphics[width=1.0\linewidth]{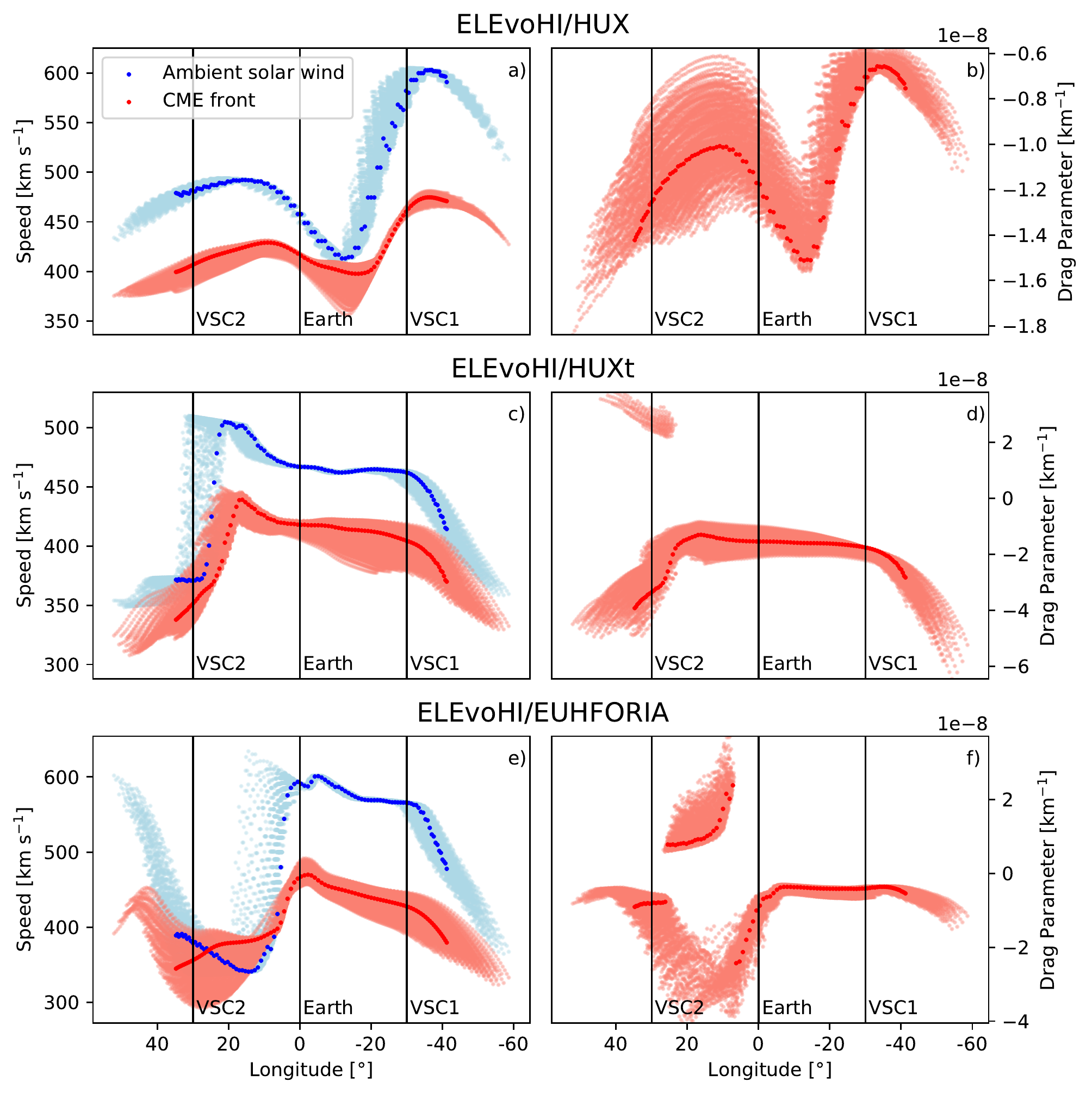}
\caption{CME front parameters at the modeled arrival time. Panels a), c), e): Ambient solar wind speed (blue) and CME speed of the deformed front (red) of each individual ensemble member and the different ambient solar wind models. Panels b), d), f): Drag parameter for each ensemble member and ambient solar wind models. The dark colors represent the values for one individual ensemble member.} \label{fig:speedDist}
\end{figure}

We are further interested in the CME frontal speed for the three different ambient solar wind models. We therefore plot the speed of the ambient solar wind and the frontal speed at the time when the front is estimated to arrive at Earth (see Figure \ref{fig:speedDist}) with the drag parameter for the ambient solar wind models.
The CME frontal speed (red in the left panels in Figure \ref{fig:speedDist}) resembles the shape of the CME front. Also the drag parameter seems to show the same behavior as the ambient solar wind. The most striking feature is that the sign of $\gamma$ changes for different longitudes. As mentioned before, we define a negative sign of $\gamma$ to indicate an acceleration while a positive sign of $\gamma$ leads to a deceleration for this certain part of the CME front. When comparing the left and the right panels in Figure \ref{fig:speedDist} it is obvious that only such ensemble members show a change in sign of $\gamma$ for which the ambient solar wind speed is lower than the CME frontal speed of this part. This is most pronounced for the EUHFORIA based model results. 

\begin{figure}[htbp]
\centering
\includegraphics[width=1\linewidth]{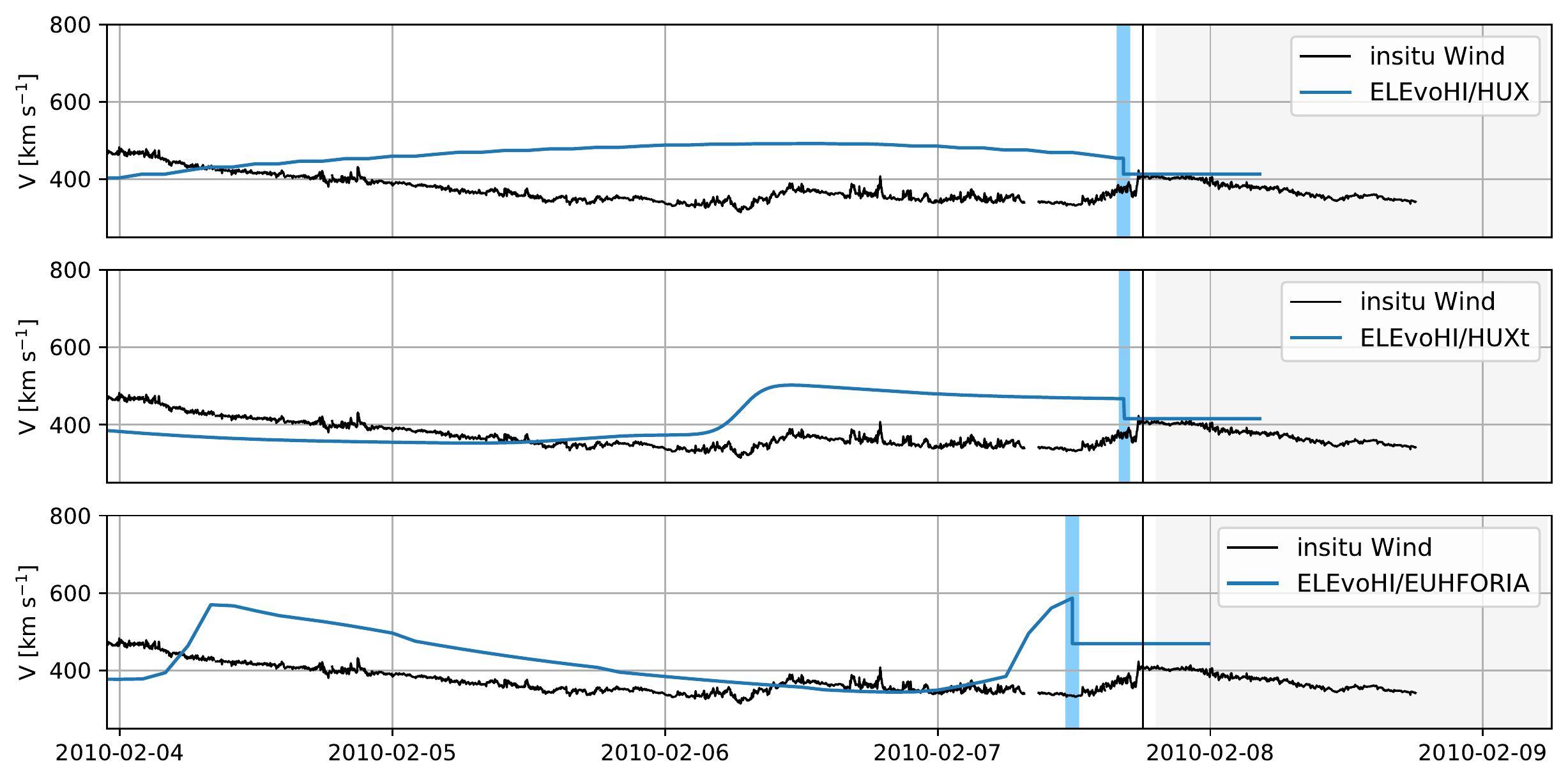}
\caption{Solar wind speed profiles for Earth direction. The black line is the in situ speed, while the blue line represents the modeled solar wind speed. The vertical solid black line indicates the in situ arrival and the vertical dashed black line is the start of the magnetic flux rope. The blue vertical bar indicates the modeled arrival time with its uncertainty. Up to that time, the speed is taken from the ambient solar wind models, afterwards the speed is set to the calculated CME arrival speed. From top to bottom the results for ELEvoHI/HUX, ELEvoHI/HUXt, and ELEvoHI/EUHFORIA are shown.} \label{fig:SpeedProfile}
\end{figure}

The actual in situ arrival speed is given by $406\pm2$~km~s$^{-1}$. The modeled arrival speeds are $413\pm3$~km~s$^{-1}$ for ELEvoHI/HUX, $416\pm3$~km~s$^{-1}$ for ELEvoHI/HUXt and $469\pm7$~km~s$^{-1}$ for ELEvoHI/EUHFORIA, where the speed corresponds to the median of all the ensemble members and the uncertainty is given by the standard deviation. The high overestimation of the calculated arrival speed also explains the early arrival when using EUHFORIA speed maps. However, the deformable front provides better speed results than for the original version of ELEvoHI. The modeled arrival speeds for the elliptical front are $474\pm7$~km~s$^{-1}$ for ELEvoHI/HUX, $461\pm4$~km~s$^{-1}$ for ELEvoHI/HUXt and $492\pm12$~km~s$^{-1}$ for ELEvoHI/EUHFORIA.

In Figure \ref{fig:SpeedProfile} the speed profiles for the three ambient solar wind models in comparison to the in situ wind speed are shown. We indicate the modeled arrival time by the vertical blue bar, where the uncertainty is given by the standard deviation of all the ensemble members that are estimated to hit Earth. Before the modeled arrival time the solar wind speed is taken from the ambient solar wind models. After that time, the calculated CME arrival speed is plotted for half a day. We can see that HUX already overestimates the ambient solar wind speed about three days prior to the in situ arrival time. The HUXt model seems to correctly model a small speed enhancement at around February 6, 2010 04:00 UT. However from this time on, also HUXt overestimates the in situ speed. EUHFORIA shows a good agreement with the in situ speed but seems to be shifted roughly by one day. Also the speed after about February 7, 2010 06:00 UT is highly overestimated. From Figure \ref{fig:SpeedProfile} we see that all of the models provide ambient solar wind speeds that are too fast compared to the measurements. The figure further shows that the modeled arrival time and speed match the actual in situ arrival quite well for ELEvoHI/HUX and ELEvoHI/HUXt. For ELEvoHI/EUHFORIA the arrival is estimated too early and too fast. Interestingly, the modeled speed profiles behave contrary to the measured speed profiles. The in situ speed is slightly slower before the defined CME arrival time and increases when the CME passes the Wind spacecraft. The modeled wind profiles, however, show a decrease of solar wind speed at arrival.

\subsection{Shifting Earth} \label{subsec:shiftingEarth}
A different approach to get an estimate of the uncertainty of the modeled CME arrival time is to artificially shift Earth position. This means that we do not consider longitude 0\textdegree\ to be the location of Earth (see Figure \ref{fig:speedDist}) but shift Earth to $\pm$10\textdegree. By doing so, we get a calculated arrival time for +10\textdegree\ of February 07, 2010 16:07 UT $\pm$ 1.8 hours and for -10\textdegree\ February 07, 2010 18:07 UT $\pm$ 2.3 hours for ELEvoHI/HUX. The modeled arrival time based on ELEvoHI/HUXt gives February 07, 2010 16:42 UT $\pm$ 2.0 hours for +10\textdegree\ and February 07, 2010 16:42 UT $\pm$ 1.8 for -10\textdegree\ and ELEvoHI/EUHFORIA models an arrival at February 07, 2010 21:07 UT $\pm$ 2.6 hours for +10\textdegree\ and February 07, 2010 12:07 UT $\pm$ 1.6 for -10\textdegree. The calculated arrival times for ELEvoHI/HUX differ by 2 hours, with the -10\textdegree\ being almost spot on regarding the in situ arrival time. ELEvoHI/HUXt provides exactly the same modeled arrival time, which is still about 1.5 hours too early. A quite different result is found ELEvoHI/EUHFORIA. For this ambient solar wind model we obtain the largest differences of 9 hours. This result is not surprising when having a look at Figure \ref{fig:speedDist}. It can be seen that the modeled speed is much slower for the ELEvoHI/EUHFORIA ambient solar wind speed at +10\textdegree\, leading to a much later calculated  arrival time.

\section{Discussion and Conclusions} \label{Sec:Summary}

In this study we present a new method for a deformable front based on ELEvoHI. The original version of ELEvoHI accounts for the drag exerted by the ambient solar wind. However, the kinematic of a CME obtained by DBM fitting is assumed only for the apex of the CME. Furthermore, the drag parameter and the ambient solar wind speed are assumed to be constant during the entire propagation in the heliosphere. With the new approach of a deformable front, ELEvoHI is able to adapt to the ambient solar wind conditions not only at the apex, but along the whole CME front. The new version of ELEvoHI can handle three different ambient solar wind models: HUX, HUXt, and EUHFORIA.

We test the deformable front by studying a CME first observed in STEREO-A/HI on February 3, 2010 14:49 UT, which has a defined in situ arrival time on Februray 7, 2010 18:04 UT and a measured speed of 406~$\pm$2~km~s$^{-1}$. In addition to Earth direction, we also model the arrival times for two additional locations in the heliosphere, defined to be $\pm$30\textdegree\ East and West of Earth (VSC1 and VSC2). We compare the calculated arrival times based on the three different ambient solar wind models for the original implementation of ELEvoHI, i.e.\ the elliptical front. For Earth direction the modeled arrival times differ at maximum 2.5~hours. However, the best model result (ELEvoHI/HUXt) is still 6~hours too early with respect to the in situ arrival time. For VSC1 and VSC2 the model results differ at maximum 3.5 and 3~hours, respectively. Considering the deformable front, we find quite different results. ELEvoHI/HUX and ELEvoHI/HUXt model an almost identical arrival time (less than 2~hours too early with respect to the in situ arrival time), while ELEvoHI/EUHFORIA models the arrival time 4.5~hours earlier compared the other two ambient solar wind models. The differences are even bigger when comparing the arrival times at the virtual spacecraft. At VSC1 the calculated arrival times differ up to more than 8.5~hours, while for VSC2 the differences reach even more than 10~hours for the three ambient solar wind models. For this case study, the modeled arrival times at Earth with the deformable front provide better results (at least 2.2~hours and 23~km~s$^{-1}$ for ELEvoHI/EUFHORIA) than the elliptical front for all the three ambient solar wind models used.

With this new approach it is further possible to get an estimate of the CME mass based on DBM fitting to the heliospheric imager data and an estimate of the cross-sectional area. For this event it could be shown that the CME mass is close to the results purely based on coronagraph images, which is in agreement with \cite{Amerstorfer2018}, who applied ELEvoHI to a halo CME event and found similar results.

Additionally, all the parameters important for the propagation of the CME front in the heliosphere can now be studied in detail at each time and location (see Figure~\ref{fig:Behavior} for three distinct directions). The solar wind density, $n$, decreases with increasing distance to the Sun, which also leads to a decreasing drag parameter, $\gamma$. The CME continually adjusts to the ambient solar wind speed the further out it propagates in the heliosphere. Both, the modeled CME frontal speed and drag parameter, resemble the CME shape quite well (see Figure~\ref{fig:speedDist}). Also, most parts of the CME front show acceleration while some parts (especially for ELEvoHI/EUHFORIA) are decelerated.

For the CME treated in this case study, we obtain almost perfect arrival speeds for ELEvoHI/HUX and ELEvoHI/HUXt, while it is overestimated by about 60~km~s$^{-1}$ by ELEvoHI/EUHFORIA. Interestingly, all of the ambient solar wind models overestimate the solar wind speed about one day before the actual in situ arrival. This leads to a modeled speed profile that is contrary to the measured speed profile. In the data we see an increase in solar wind speed up to the in situ arrival time, while in the modeled profile the speed drops at the calculated arrival time.

We also study the arrival time uncertainties by shifting Earth to different locations (e.g.\ $\pm$10\textdegree, see Section \ref{subsec:shiftingEarth}).
We find that for ambient solar wind models, which exhibit more structured ambient solar wind conditions, the uncertainties in the arrival time increases. In the case of ELEvoHI/EUHFORIA the modeled arrival times differ up to more than 9~hours. This is again in the range of our current forecast capabilities. It also shows that ELEvoHI is highly dependent on accurate ambient solar wind models but those are known to have substantial inherent uncertainties by themselves.

In this study we consider the CME arrival times and speed only in the ecliptic plane, even though the ambient solar wind and CMEs are 3D phenomena. Therefore, we do not provide any uncertainties regarding the modeled CME arrival depending on the latitude. However, we expect the uncertainties to be in the same range as when shifting the Earth to different longitudes.

In the previous version of ELEvoHI the CMEs are treated as coherent structures, meaning that the frontal shape, once defined, does not change during propagation. Hence, it assumes that the internal magnetic field and the associated magnetic tension force prevents the CME from deformation. \cite{Owens2017Nat} showed that at about 0.3~AU the majority of CMEs do not behave as coherent structures anymore. As a consequence the different flanks of a CME are effectively independent from each other, while neighbouring parts of the CME front are most likely to experience magnetic tension. In the current implementation of ELEvoHI 2.0 each point of the CME front propagates individually, i.e. no structural coherence is given. However, the results obtained in this study indicate that the CME fronts do not show discontinuities for the three ambient solar wind models used. The reason is mainly due to the relatively small change of ambient solar wind speed from one longitude to the next.

Recent studies \citep[e.g.][]{Barnard2017,Kay2021,Wang2016,Zhuang2017} have shown the importance of deformation, but also deflection and expansion of CMEs to obtain more accurate CME arrival time predictions for drag-based models. Associated to that, an evaluation of the drag parameter along the whole CME front is required. Also CME-CME interaction is essential for arrival time prediction. However, such interactions are not incorporated in the current version of ELEvoHI 2.0. A preceding CME leads to a preconditioning of the ambient solar wind \citep[e.g.][]{Temmer2017}, which is so far not implemented in the solar wind models used by our model.
This study is only a first step to a better understanding of the CME propagation behavior in the heliosphere. Future work will include a broader test based on a larger sample of events to detect and constrain the important factors influencing CME arrival predictions.

\section{Data Sources} \label{sec:DataSources}

\noindent \textbf{Data} \\

\noindent STEREO/HI: \url{https://www.ukssdc.ac.uk/solar/stereo/data.html} \\
\noindent STEREO/COR2: \url{https://stereo-ssc.nascom.nasa.gov/data/} \\
\noindent HELCATS: \url{https://www.helcats-fp7.eu}\\
\noindent ICMECAT: \url{https://doi.org/10.6084/m9.figshare.6356420}

\noindent \textbf{Model} \\

\noindent ELEvoHI 2.0 is available at \url{https://doi.org/10.5281/zenodo.5045415}

\noindent \textbf{Results} \\

\noindent The visualization of each model result, i.e.\ movies and figures, as well as the results from the ambient solar wind models can be downloaded from \url{https://doi.org/10.6084/m9.figshare.14923032.v1}.

\noindent \textbf{Software} \\

\noindent IDL\textsuperscript{TM} Version 8.4 \\
\noindent Python 3.7.6 \\
\noindent SATPLOT: \url{https://hesperia.gsfc.nasa.gov/ssw/stereo/secchi/idl/jpl/satplot/SATPLOT\_User\_Guide.pdf}

%
%
%
%
%
%
%
%

\acknowledgments
J.H., T.A., M.A.R., A.J.W., C.M., M.B. and U.V.A. thank the Austrian Science Fund (FWF): P31265-N27, P31659-N27, P31521-N27. STEREO/HI was developed by a consortium comprising Rutherford Appleton Laboratory, and University of Birmingham (UK), Centre Spatiale de Liége (Belgium), and the Naval Research Laboratory (USA). The authors acknowledge the UK Solar System Data Centre for provision of the STEREO/HI data. The STEREO/SECCHI data used here are produced by an international consortium of the Naval Research Laboratory (USA), Lockheed Martin Solar and Astrophysics Laboratory (USA), NASA Goddard Space Flight Center (USA), Rutherford Appleton Laboratory (UK), University of Birmingham (UK), Max-Planck-Institut für Sonnensystemforschung (Germany), Centre Spatiale de Liège (Belgium), Institut d'Optique Théorique et Appliqué (France), and Institut d'Astrophysique Spatiale (France).


%
\bibliography{juh_bib}

\begin{thebibliography}{}

\bibitem [\protect \citeauthoryear {%
{Altschuler}%
\ \BBA {} {Newkirk}%
}{%
{Altschuler}%
\ \BBA {} {Newkirk}%
}{%
{\protect \APACyear {1969}}%
}]{%
altschuler69}
\APACinsertmetastar {%
altschuler69}%
\begin{APACrefauthors}%
{Altschuler}, M\BPBI D.%
\BCBT {}\ \BBA {} {Newkirk}, G.%
\end{APACrefauthors}%
\unskip\
\newblock
\APACrefYearMonthDay{1969}{{\APACmonth{09}}}{}.
\newblock
{\BBOQ}\APACrefatitle {{Magnetic Fields and the Structure of the Solar Corona.
  I: Methods of Calculating Coronal Fields}} {{Magnetic Fields and the
  Structure of the Solar Corona. I: Methods of Calculating Coronal
  Fields}}.{\BBCQ}
\newblock
\APACjournalVolNumPages{\solphys}{9}{}{131-149}.
\newblock
\begin{APACrefDOI} \doi{10.1007/BF00145734} \end{APACrefDOI}
\PrintBackRefs{\CurrentBib}

\bibitem [\protect \citeauthoryear {%
Amerstorfer%
\ \protect \BOthers {.}}{%
Amerstorfer%
\ \protect \BOthers {.}}{%
{\protect \APACyear {2021}}%
}]{%
Amerstorfer2021}
\APACinsertmetastar {%
Amerstorfer2021}%
\begin{APACrefauthors}%
Amerstorfer, T.%
, Hinterreiter, J.%
, Reiss, M\BPBI A.%
, Möstl, C.%
, Davies, J\BPBI A.%
, Bailey, R\BPBI L.%
\BDBL {}Harrison, R\BPBI A.%
\end{APACrefauthors}%
\unskip\
\newblock
\APACrefYearMonthDay{2021}{}{}.
\newblock
{\BBOQ}\APACrefatitle {Evaluation of CME Arrival Prediction Using Ensemble
  Modeling Based on Heliospheric Imaging Observations} {Evaluation of cme
  arrival prediction using ensemble modeling based on heliospheric imaging
  observations}.{\BBCQ}
\newblock
\APACjournalVolNumPages{Space Weather}{19}{1}{e2020SW002553}.
\newblock
\begin{APACrefURL}
  \url{https://agupubs.onlinelibrary.wiley.com/doi/abs/10.1029/2020SW002553}
  \end{APACrefURL}
\newblock
\APACrefnote{e2020SW002553 10.1029/2020SW002553}
\newblock
\begin{APACrefDOI} \doi{https://doi.org/10.1029/2020SW002553} \end{APACrefDOI}
\PrintBackRefs{\CurrentBib}

\bibitem [\protect \citeauthoryear {%
{Amerstorfer}%
\ \protect \BOthers {.}}{%
{Amerstorfer}%
\ \protect \BOthers {.}}{%
{\protect \APACyear {2018}}%
}]{%
Amerstorfer2018}
\APACinsertmetastar {%
Amerstorfer2018}%
\begin{APACrefauthors}%
{Amerstorfer}, T.%
, {M{\"o}stl}, C.%
, {Hess}, P.%
, {Temmer}, M.%
, {Mays}, M\BPBI L.%
, {Reiss}, M\BPBI A.%
\BDBL {}{Bourdin}, P\BPBI A.%
\end{APACrefauthors}%
\unskip\
\newblock
\APACrefYearMonthDay{2018}{{\APACmonth{07}}}{}.
\newblock
{\BBOQ}\APACrefatitle {{Ensemble Prediction of a Halo Coronal Mass Ejection
  Using Heliospheric Imagers}} {{Ensemble Prediction of a Halo Coronal Mass
  Ejection Using Heliospheric Imagers}}.{\BBCQ}
\newblock
\APACjournalVolNumPages{Space Weather}{16}{7}{784-801}.
\newblock
\begin{APACrefDOI} \doi{10.1029/2017SW001786} \end{APACrefDOI}
\PrintBackRefs{\CurrentBib}

\bibitem [\protect \citeauthoryear {%
{Arge}%
, {Odstrcil}%
, {Pizzo}%
\BCBL {}\ \BBA {} {Mayer}%
}{%
{Arge}%
\ \protect \BOthers {.}}{%
{\protect \APACyear {2003}}%
}]{%
arge03}
\APACinsertmetastar {%
arge03}%
\begin{APACrefauthors}%
{Arge}, C\BPBI N.%
, {Odstrcil}, D.%
, {Pizzo}, V\BPBI J.%
\BCBL {}\ \BBA {} {Mayer}, L\BPBI R.%
\end{APACrefauthors}%
\unskip\
\newblock
\APACrefYearMonthDay{2003}{{\APACmonth{09}}}{}.
\newblock
{\BBOQ}\APACrefatitle {{Improved Method for Specifying Solar Wind Speed Near
  the Sun}} {{Improved Method for Specifying Solar Wind Speed Near the
  Sun}}.{\BBCQ}
\newblock
\BIn{} M.~{Velli}, R.~{Bruno}, F.~{Malara}\BCBL {}\ \BBA {} B.~{Bucci}\
  (\BEDS), \APACrefbtitle {Solar Wind Ten} {Solar wind ten}\ (\BVOL~679,
  \BPG~190-193).
\newblock
\begin{APACrefDOI} \doi{10.1063/1.1618574} \end{APACrefDOI}
\PrintBackRefs{\CurrentBib}

\bibitem [\protect \citeauthoryear {%
Barnard%
, Owens%
, Scott%
\BCBL {}\ \BBA {} de Koning%
}{%
Barnard%
\ \protect \BOthers {.}}{%
{\protect \APACyear {2020}}%
}]{%
Barnard2020}
\APACinsertmetastar {%
Barnard2020}%
\begin{APACrefauthors}%
Barnard, L.%
, Owens, M\BPBI J.%
, Scott, C\BPBI J.%
\BCBL {}\ \BBA {} de Koning, C\BPBI A.%
\end{APACrefauthors}%
\unskip\
\newblock
\APACrefYearMonthDay{2020}{}{}.
\newblock
{\BBOQ}\APACrefatitle {Ensemble CME Modeling Constrained by Heliospheric Imager
  Observations} {Ensemble cme modeling constrained by heliospheric imager
  observations}.{\BBCQ}
\newblock
\APACjournalVolNumPages{AGU Advances}{1}{3}{e2020AV000214}.
\newblock
\begin{APACrefURL}
  \url{https://agupubs.onlinelibrary.wiley.com/doi/abs/10.1029/2020AV000214}
  \end{APACrefURL}
\newblock
\APACrefnote{e2020AV000214 10.1029/2020AV000214}
\newblock
\begin{APACrefDOI} \doi{10.1029/2020AV000214} \end{APACrefDOI}
\PrintBackRefs{\CurrentBib}

\bibitem [\protect \citeauthoryear {%
{Barnard}%
\ \protect \BOthers {.}}{%
{Barnard}%
\ \protect \BOthers {.}}{%
{\protect \APACyear {2017}}%
}]{%
Barnard2017}
\APACinsertmetastar {%
Barnard2017}%
\begin{APACrefauthors}%
{Barnard}, L\BPBI A.%
, {de Koning}, C\BPBI A.%
, {Scott}, C\BPBI J.%
, {Owens}, M\BPBI J.%
, {Wilkinson}, J.%
\BCBL {}\ \BBA {} {Davies}, J\BPBI A.%
\end{APACrefauthors}%
\unskip\
\newblock
\APACrefYearMonthDay{2017}{{\APACmonth{06}}}{}.
\newblock
{\BBOQ}\APACrefatitle {{Testing the current paradigm for space weather
  prediction with heliospheric imagers}} {{Testing the current paradigm for
  space weather prediction with heliospheric imagers}}.{\BBCQ}
\newblock
\APACjournalVolNumPages{Space Weather}{15}{6}{782-803}.
\newblock
\begin{APACrefDOI} \doi{10.1002/2017SW001609} \end{APACrefDOI}
\PrintBackRefs{\CurrentBib}

\bibitem [\protect \citeauthoryear {%
{Bein}%
, {Temmer}%
, {Vourlidas}%
, {Veronig}%
\BCBL {}\ \BBA {} {Utz}%
}{%
{Bein}%
\ \protect \BOthers {.}}{%
{\protect \APACyear {2013}}%
}]{%
Bein2013}
\APACinsertmetastar {%
Bein2013}%
\begin{APACrefauthors}%
{Bein}, B\BPBI M.%
, {Temmer}, M.%
, {Vourlidas}, A.%
, {Veronig}, A\BPBI M.%
\BCBL {}\ \BBA {} {Utz}, D.%
\end{APACrefauthors}%
\unskip\
\newblock
\APACrefYearMonthDay{2013}{{\APACmonth{05}}}{}.
\newblock
{\BBOQ}\APACrefatitle {{The Height Evolution of the ``True'' Coronal Mass
  Ejection Mass derived from STEREO COR1 and COR2 Observations}} {{The Height
  Evolution of the ``True'' Coronal Mass Ejection Mass derived from STEREO COR1
  and COR2 Observations}}.{\BBCQ}
\newblock
\APACjournalVolNumPages{\apj}{768}{1}{31}.
\newblock
\begin{APACrefDOI} \doi{10.1088/0004-637X/768/1/31} \end{APACrefDOI}
\PrintBackRefs{\CurrentBib}

\bibitem [\protect \citeauthoryear {%
{Billings}%
}{%
{Billings}%
}{%
{\protect \APACyear {1966}}%
}]{%
Billings1966}
\APACinsertmetastar {%
Billings1966}%
\begin{APACrefauthors}%
{Billings}, D\BPBI E.%
\end{APACrefauthors}%
\unskip\
\newblock
\APACrefYear{1966}.
\newblock
\APACrefbtitle {{A guide to the solar corona}} {{A guide to the solar corona}}.
\PrintBackRefs{\CurrentBib}

\bibitem [\protect \citeauthoryear {%
{Cannon}%
}{%
{Cannon}%
}{%
{\protect \APACyear {2013}}%
}]{%
Cannon2013}
\APACinsertmetastar {%
Cannon2013}%
\begin{APACrefauthors}%
{Cannon}, P\BPBI S.%
\end{APACrefauthors}%
\unskip\
\newblock
\APACrefYearMonthDay{2013}{{\APACmonth{04}}}{}.
\newblock
{\BBOQ}\APACrefatitle {{Extreme Space Weather{\textemdash}A Report Published by
  the UK Royal Academy of Engineering}} {{Extreme Space Weather{\textemdash}A
  Report Published by the UK Royal Academy of Engineering}}.{\BBCQ}
\newblock
\APACjournalVolNumPages{Space Weather}{11}{4}{138-139}.
\newblock
\begin{APACrefDOI} \doi{10.1002/swe.20032} \end{APACrefDOI}
\PrintBackRefs{\CurrentBib}

\bibitem [\protect \citeauthoryear {%
{Cargill}%
}{%
{Cargill}%
}{%
{\protect \APACyear {2004}}%
}]{%
Cargill2004}
\APACinsertmetastar {%
Cargill2004}%
\begin{APACrefauthors}%
{Cargill}, P\BPBI J.%
\end{APACrefauthors}%
\unskip\
\newblock
\APACrefYearMonthDay{2004}{{\APACmonth{05}}}{}.
\newblock
{\BBOQ}\APACrefatitle {{On the Aerodynamic Drag Force Acting on Interplanetary
  Coronal Mass Ejections}} {{On the Aerodynamic Drag Force Acting on
  Interplanetary Coronal Mass Ejections}}.{\BBCQ}
\newblock
\APACjournalVolNumPages{\solphys}{221}{1}{135-149}.
\newblock
\begin{APACrefDOI} \doi{10.1023/B:SOLA.0000033366.10725.a2} \end{APACrefDOI}
\PrintBackRefs{\CurrentBib}

\bibitem [\protect \citeauthoryear {%
{Colaninno}%
\ \BBA {} {Vourlidas}%
}{%
{Colaninno}%
\ \BBA {} {Vourlidas}%
}{%
{\protect \APACyear {2009}}%
}]{%
Colaninno2009}
\APACinsertmetastar {%
Colaninno2009}%
\begin{APACrefauthors}%
{Colaninno}, R\BPBI C.%
\BCBT {}\ \BBA {} {Vourlidas}, A.%
\end{APACrefauthors}%
\unskip\
\newblock
\APACrefYearMonthDay{2009}{{\APACmonth{06}}}{}.
\newblock
{\BBOQ}\APACrefatitle {{First Determination of the True Mass of Coronal Mass
  Ejections: A Novel Approach to Using the Two STEREO Viewpoints}} {{First
  Determination of the True Mass of Coronal Mass Ejections: A Novel Approach to
  Using the Two STEREO Viewpoints}}.{\BBCQ}
\newblock
\APACjournalVolNumPages{\apj}{698}{1}{852-858}.
\newblock
\begin{APACrefDOI} \doi{10.1088/0004-637X/698/1/852} \end{APACrefDOI}
\PrintBackRefs{\CurrentBib}

\bibitem [\protect \citeauthoryear {%
{Davies}%
\ \protect \BOthers {.}}{%
{Davies}%
\ \protect \BOthers {.}}{%
{\protect \APACyear {2012}}%
}]{%
Davies2012}
\APACinsertmetastar {%
Davies2012}%
\begin{APACrefauthors}%
{Davies}, J\BPBI A.%
, {Harrison}, R\BPBI A.%
, {Perry}, C\BPBI H.%
, {M{\"o}stl}, C.%
, {Lugaz}, N.%
, {Rollett}, T.%
\BDBL {}{Savani}, N\BPBI P.%
\end{APACrefauthors}%
\unskip\
\newblock
\APACrefYearMonthDay{2012}{{\APACmonth{05}}}{}.
\newblock
{\BBOQ}\APACrefatitle {{A Self-similar Expansion Model for Use in Solar Wind
  Transient Propagation Studies}} {{A Self-similar Expansion Model for Use in
  Solar Wind Transient Propagation Studies}}.{\BBCQ}
\newblock
\APACjournalVolNumPages{\apj}{750}{1}{23}.
\newblock
\begin{APACrefDOI} \doi{10.1088/0004-637X/750/1/23} \end{APACrefDOI}
\PrintBackRefs{\CurrentBib}

\bibitem [\protect \citeauthoryear {%
{Davies}%
\ \protect \BOthers {.}}{%
{Davies}%
\ \protect \BOthers {.}}{%
{\protect \APACyear {2009}}%
}]{%
Davies2009}
\APACinsertmetastar {%
Davies2009}%
\begin{APACrefauthors}%
{Davies}, J\BPBI A.%
, {Harrison}, R\BPBI A.%
, {Rouillard}, A\BPBI P.%
, {Sheeley}, N\BPBI R.%
, {Perry}, C\BPBI H.%
, {Bewsher}, D.%
\BDBL {}{Brown}, D\BPBI S.%
\end{APACrefauthors}%
\unskip\
\newblock
\APACrefYearMonthDay{2009}{{\APACmonth{01}}}{}.
\newblock
{\BBOQ}\APACrefatitle {{A synoptic view of solar transient evolution in the
  inner heliosphere using the Heliospheric Imagers on STEREO}} {{A synoptic
  view of solar transient evolution in the inner heliosphere using the
  Heliospheric Imagers on STEREO}}.{\BBCQ}
\newblock
\APACjournalVolNumPages{\grl}{36}{2}{L02102}.
\newblock
\begin{APACrefDOI} \doi{10.1029/2008GL036182} \end{APACrefDOI}
\PrintBackRefs{\CurrentBib}

\bibitem [\protect \citeauthoryear {%
{de Koning}%
}{%
{de Koning}%
}{%
{\protect \APACyear {2017}}%
}]{%
DeKoning2017}
\APACinsertmetastar {%
DeKoning2017}%
\begin{APACrefauthors}%
{de Koning}, C\BPBI A.%
\end{APACrefauthors}%
\unskip\
\newblock
\APACrefYearMonthDay{2017}{{\APACmonth{07}}}{}.
\newblock
{\BBOQ}\APACrefatitle {{Lessons Learned from the Three-view Determination of
  CME Mass}} {{Lessons Learned from the Three-view Determination of CME
  Mass}}.{\BBCQ}
\newblock
\APACjournalVolNumPages{\apj}{844}{1}{61}.
\newblock
\begin{APACrefDOI} \doi{10.3847/1538-4357/aa7a09} \end{APACrefDOI}
\PrintBackRefs{\CurrentBib}

\bibitem [\protect \citeauthoryear {%
{Dumbovi{\'c}}%
\ \protect \BOthers {.}}{%
{Dumbovi{\'c}}%
\ \protect \BOthers {.}}{%
{\protect \APACyear {2018}}%
}]{%
Dumbovic2018}
\APACinsertmetastar {%
Dumbovic2018}%
\begin{APACrefauthors}%
{Dumbovi{\'c}}, M.%
, {{\v{C}}alogovi{\'c}}, J.%
, {Vr{\v{s}}nak}, B.%
, {Temmer}, M.%
, {Mays}, M\BPBI L.%
, {Veronig}, A.%
\BCBL {}\ \BBA {} {Piantschitsch}, I.%
\end{APACrefauthors}%
\unskip\
\newblock
\APACrefYearMonthDay{2018}{{\APACmonth{02}}}{}.
\newblock
{\BBOQ}\APACrefatitle {{The Drag-based Ensemble Model (DBEM) for Coronal Mass
  Ejection Propagation}} {{The Drag-based Ensemble Model (DBEM) for Coronal
  Mass Ejection Propagation}}.{\BBCQ}
\newblock
\APACjournalVolNumPages{\apj}{854}{2}{180}.
\newblock
\begin{APACrefDOI} \doi{10.3847/1538-4357/aaaa66} \end{APACrefDOI}
\PrintBackRefs{\CurrentBib}

\bibitem [\protect \citeauthoryear {%
{Eyles}%
\ \protect \BOthers {.}}{%
{Eyles}%
\ \protect \BOthers {.}}{%
{\protect \APACyear {2009}}%
}]{%
Eyles2009}
\APACinsertmetastar {%
Eyles2009}%
\begin{APACrefauthors}%
{Eyles}, C\BPBI J.%
, {Harrison}, R\BPBI A.%
, {Davis}, C\BPBI J.%
, {Waltham}, N\BPBI R.%
, {Shaughnessy}, B\BPBI M.%
, {Mapson-Menard}, H\BPBI C\BPBI A.%
\BDBL {}{Rochus}, P.%
\end{APACrefauthors}%
\unskip\
\newblock
\APACrefYearMonthDay{2009}{{\APACmonth{02}}}{}.
\newblock
{\BBOQ}\APACrefatitle {{The Heliospheric Imagers Onboard the STEREO Mission}}
  {{The Heliospheric Imagers Onboard the STEREO Mission}}.{\BBCQ}
\newblock
\APACjournalVolNumPages{\solphys}{254}{2}{387-445}.
\newblock
\begin{APACrefDOI} \doi{10.1007/s11207-008-9299-0} \end{APACrefDOI}
\PrintBackRefs{\CurrentBib}

\bibitem [\protect \citeauthoryear {%
{Eyni}%
\ \BBA {} {Steinitz}%
}{%
{Eyni}%
\ \BBA {} {Steinitz}%
}{%
{\protect \APACyear {1980}}%
}]{%
EyniSteinitz1980}
\APACinsertmetastar {%
EyniSteinitz1980}%
\begin{APACrefauthors}%
{Eyni}, M.%
\BCBT {}\ \BBA {} {Steinitz}, R.%
\end{APACrefauthors}%
\unskip\
\newblock
\APACrefYearMonthDay{1980}{{\APACmonth{01}}}{}.
\newblock
{\BBOQ}\APACrefatitle {{An empirical relation between density, flow velocity
  and heliocentric distance in the solar wind}} {{An empirical relation between
  density, flow velocity and heliocentric distance in the solar wind}}.{\BBCQ}
\newblock
\BIn{} M.~{Dryer}\ \BBA {} E.~{Tandberg-Hanssen}\ (\BEDS), \APACrefbtitle
  {Solar and Interplanetary Dynamics} {Solar and interplanetary dynamics}\
  (\BVOL~91, \BPG~147-149).
\PrintBackRefs{\CurrentBib}

\bibitem [\protect \citeauthoryear {%
{Gopalswamy}%
\ \protect \BOthers {.}}{%
{Gopalswamy}%
\ \protect \BOthers {.}}{%
{\protect \APACyear {2000}}%
}]{%
Gopalswamy2000}
\APACinsertmetastar {%
Gopalswamy2000}%
\begin{APACrefauthors}%
{Gopalswamy}, N.%
, {Lara}, A.%
, {Lepping}, R\BPBI P.%
, {Kaiser}, M\BPBI L.%
, {Berdichevsky}, D.%
\BCBL {}\ \BBA {} {St. Cyr}, O\BPBI C.%
\end{APACrefauthors}%
\unskip\
\newblock
\APACrefYearMonthDay{2000}{{\APACmonth{01}}}{}.
\newblock
{\BBOQ}\APACrefatitle {{Interplanetary acceleration of coronal mass ejections}}
  {{Interplanetary acceleration of coronal mass ejections}}.{\BBCQ}
\newblock
\APACjournalVolNumPages{\grl}{27}{2}{145-148}.
\newblock
\begin{APACrefDOI} \doi{10.1029/1999GL003639} \end{APACrefDOI}
\PrintBackRefs{\CurrentBib}

\bibitem [\protect \citeauthoryear {%
{Gosling}%
, {Bame}%
, {McComas}%
\BCBL {}\ \BBA {} {Phillips}%
}{%
{Gosling}%
\ \protect \BOthers {.}}{%
{\protect \APACyear {1990}}%
}]{%
Gosling1990}
\APACinsertmetastar {%
Gosling1990}%
\begin{APACrefauthors}%
{Gosling}, J\BPBI T.%
, {Bame}, S\BPBI J.%
, {McComas}, D\BPBI J.%
\BCBL {}\ \BBA {} {Phillips}, J\BPBI L.%
\end{APACrefauthors}%
\unskip\
\newblock
\APACrefYearMonthDay{1990}{{\APACmonth{06}}}{}.
\newblock
{\BBOQ}\APACrefatitle {{Coronal mass ejections and large geomagnetic storms}}
  {{Coronal mass ejections and large geomagnetic storms}}.{\BBCQ}
\newblock
\APACjournalVolNumPages{\grl}{17}{7}{901-904}.
\newblock
\begin{APACrefDOI} \doi{10.1029/GL017i007p00901} \end{APACrefDOI}
\PrintBackRefs{\CurrentBib}

\bibitem [\protect \citeauthoryear {%
{Gui}%
\ \protect \BOthers {.}}{%
{Gui}%
\ \protect \BOthers {.}}{%
{\protect \APACyear {2011}}%
}]{%
GUI2011}
\APACinsertmetastar {%
GUI2011}%
\begin{APACrefauthors}%
{Gui}, B.%
, {Shen}, C.%
, {Wang}, Y.%
, {Ye}, P.%
, {Liu}, J.%
, {Wang}, S.%
\BCBL {}\ \BBA {} {Zhao}, X.%
\end{APACrefauthors}%
\unskip\
\newblock
\APACrefYearMonthDay{2011}{{\APACmonth{07}}}{}.
\newblock
{\BBOQ}\APACrefatitle {{Quantitative Analysis of CME Deflections in the
  Corona}} {{Quantitative Analysis of CME Deflections in the Corona}}.{\BBCQ}
\newblock
\APACjournalVolNumPages{\solphys}{271}{1-2}{111-139}.
\newblock
\begin{APACrefDOI} \doi{10.1007/s11207-011-9791-9} \end{APACrefDOI}
\PrintBackRefs{\CurrentBib}

\bibitem [\protect \citeauthoryear {%
Hinterreiter%
\ \protect \BOthers {.}}{%
Hinterreiter%
\ \protect \BOthers {.}}{%
{\protect \APACyear {2021}}%
}]{%
Hinterreiter2021}
\APACinsertmetastar {%
Hinterreiter2021}%
\begin{APACrefauthors}%
Hinterreiter, J.%
, Amerstorfer, T.%
, Reiss, M\BPBI A.%
, Möstl, C.%
, Temmer, M.%
, Bauer, M.%
\BDBL {}Owens, M\BPBI J.%
\end{APACrefauthors}%
\unskip\
\newblock
\APACrefYearMonthDay{2021}{}{}.
\newblock
{\BBOQ}\APACrefatitle {Why are ELEvoHI CME Arrival Predictions Different if
  Based on STEREO-A or STEREO-B Heliospheric Imager Observations?} {Why are
  elevohi cme arrival predictions different if based on stereo-a or stereo-b
  heliospheric imager observations?}{\BBCQ}
\newblock
\APACjournalVolNumPages{Space Weather}{19}{3}{e2020SW002674}.
\newblock
\begin{APACrefURL}
  \url{https://agupubs.onlinelibrary.wiley.com/doi/abs/10.1029/2020SW002674}
  \end{APACrefURL}
\newblock
\APACrefnote{e2020SW002674 2020SW002674}
\newblock
\begin{APACrefDOI} \doi{https://doi.org/10.1029/2020SW002674} \end{APACrefDOI}
\PrintBackRefs{\CurrentBib}

\bibitem [\protect \citeauthoryear {%
{Howard}%
\ \BBA {} {Tappin}%
}{%
{Howard}%
\ \BBA {} {Tappin}%
}{%
{\protect \APACyear {2009}}%
{\protect \APACexlab {{\protect \BCnt {1}}}}}]{%
HowardTappin2009_1}
\APACinsertmetastar {%
HowardTappin2009_1}%
\begin{APACrefauthors}%
{Howard}, T\BPBI A.%
\BCBT {}\ \BBA {} {Tappin}, S\BPBI J.%
\end{APACrefauthors}%
\unskip\
\newblock
\APACrefYearMonthDay{2009{\protect \BCnt {1}}}{{\APACmonth{10}}}{}.
\newblock
{\BBOQ}\APACrefatitle {{Interplanetary Coronal Mass Ejections Observed in the
  Heliosphere: 1. Review of Theory}} {{Interplanetary Coronal Mass Ejections
  Observed in the Heliosphere: 1. Review of Theory}}.{\BBCQ}
\newblock
\APACjournalVolNumPages{\ssr}{147}{1-2}{31-54}.
\newblock
\begin{APACrefDOI} \doi{10.1007/s11214-009-9542-5} \end{APACrefDOI}
\PrintBackRefs{\CurrentBib}

\bibitem [\protect \citeauthoryear {%
{Howard}%
\ \BBA {} {Tappin}%
}{%
{Howard}%
\ \BBA {} {Tappin}%
}{%
{\protect \APACyear {2009}}%
{\protect \APACexlab {{\protect \BCnt {2}}}}}]{%
HowardTappin2009_3}
\APACinsertmetastar {%
HowardTappin2009_3}%
\begin{APACrefauthors}%
{Howard}, T\BPBI A.%
\BCBT {}\ \BBA {} {Tappin}, S\BPBI J.%
\end{APACrefauthors}%
\unskip\
\newblock
\APACrefYearMonthDay{2009{\protect \BCnt {2}}}{{\APACmonth{10}}}{}.
\newblock
{\BBOQ}\APACrefatitle {{Interplanetary Coronal Mass Ejections Observed in the
  Heliosphere: 3. Physical Implications}} {{Interplanetary Coronal Mass
  Ejections Observed in the Heliosphere: 3. Physical Implications}}.{\BBCQ}
\newblock
\APACjournalVolNumPages{\ssr}{147}{1-2}{89-110}.
\newblock
\begin{APACrefDOI} \doi{10.1007/s11214-009-9577-7} \end{APACrefDOI}
\PrintBackRefs{\CurrentBib}

\bibitem [\protect \citeauthoryear {%
{Howard}%
\ \BBA {} {Tappin}%
}{%
{Howard}%
\ \BBA {} {Tappin}%
}{%
{\protect \APACyear {2010}}%
}]{%
HowardTappin2010}
\APACinsertmetastar {%
HowardTappin2010}%
\begin{APACrefauthors}%
{Howard}, T\BPBI A.%
\BCBT {}\ \BBA {} {Tappin}, S\BPBI J.%
\end{APACrefauthors}%
\unskip\
\newblock
\APACrefYearMonthDay{2010}{{\APACmonth{07}}}{}.
\newblock
{\BBOQ}\APACrefatitle {{Application of a new phenomenological coronal mass
  ejection model to space weather forecasting}} {{Application of a new
  phenomenological coronal mass ejection model to space weather
  forecasting}}.{\BBCQ}
\newblock
\APACjournalVolNumPages{Space Weather}{8}{7}{S07004}.
\newblock
\begin{APACrefDOI} \doi{10.1029/2009SW000531} \end{APACrefDOI}
\PrintBackRefs{\CurrentBib}

\bibitem [\protect \citeauthoryear {%
{Hundhausen}%
, {Stanger}%
\BCBL {}\ \BBA {} {Serbicki}%
}{%
{Hundhausen}%
\ \protect \BOthers {.}}{%
{\protect \APACyear {1994}}%
}]{%
Hundhausen1994}
\APACinsertmetastar {%
Hundhausen1994}%
\begin{APACrefauthors}%
{Hundhausen}, A\BPBI J.%
, {Stanger}, A\BPBI L.%
\BCBL {}\ \BBA {} {Serbicki}, S\BPBI A.%
\end{APACrefauthors}%
\unskip\
\newblock
\APACrefYearMonthDay{1994}{{\APACmonth{12}}}{}.
\newblock
{\BBOQ}\APACrefatitle {{Mass and energy contents of coronal mass ejections: SMM
  results from 1980 and 1984-1988}} {{Mass and energy contents of coronal mass
  ejections: SMM results from 1980 and 1984-1988}}.{\BBCQ}
\newblock
\BIn{} J\BPBI J.~{Hunt}\ (\BED), \APACrefbtitle {Solar Dynamic Phenomena and
  Solar Wind Consequences, the Third SOHO Workshop} {Solar dynamic phenomena
  and solar wind consequences, the third soho workshop}\ (\BVOL~373, \BPG~409).
\PrintBackRefs{\CurrentBib}

\bibitem [\protect \citeauthoryear {%
{Kay}%
, {Mays}%
\BCBL {}\ \BBA {} {Verbeke}%
}{%
{Kay}%
\ \protect \BOthers {.}}{%
{\protect \APACyear {2020}}%
}]{%
Kay2020}
\APACinsertmetastar {%
Kay2020}%
\begin{APACrefauthors}%
{Kay}, C.%
, {Mays}, M\BPBI L.%
\BCBL {}\ \BBA {} {Verbeke}, C.%
\end{APACrefauthors}%
\unskip\
\newblock
\APACrefYearMonthDay{2020}{{\APACmonth{01}}}{}.
\newblock
{\BBOQ}\APACrefatitle {{Identifying Critical Input Parameters for Improving
  Drag-Based CME Arrival Time Predictions}} {{Identifying Critical Input
  Parameters for Improving Drag-Based CME Arrival Time Predictions}}.{\BBCQ}
\newblock
\APACjournalVolNumPages{Space Weather}{18}{1}{e02382}.
\newblock
\begin{APACrefDOI} \doi{10.1029/2019SW002382} \end{APACrefDOI}
\PrintBackRefs{\CurrentBib}

\bibitem [\protect \citeauthoryear {%
{Kay}%
\ \BBA {} {Nieves-Chinchilla}%
}{%
{Kay}%
\ \BBA {} {Nieves-Chinchilla}%
}{%
{\protect \APACyear {2020}}%
}]{%
KayNievesChinchilla2020}
\APACinsertmetastar {%
KayNievesChinchilla2020}%
\begin{APACrefauthors}%
{Kay}, C.%
\BCBT {}\ \BBA {} {Nieves-Chinchilla}, T.%
\end{APACrefauthors}%
\unskip\
\newblock
\APACrefYearMonthDay{2020}{{\APACmonth{11}}}{}.
\newblock
{\BBOQ}\APACrefatitle {{Modeling Interplanetary Expansion and Deformation of
  CMEs with ANTEATR-PARADE I: Relative Contribution of Different Forces}}
  {{Modeling Interplanetary Expansion and Deformation of CMEs with
  ANTEATR-PARADE I: Relative Contribution of Different Forces}}.{\BBCQ}
\newblock
\APACjournalVolNumPages{arXiv e-prints}{}{}{arXiv:2011.06030}.
\PrintBackRefs{\CurrentBib}

\bibitem [\protect \citeauthoryear {%
{Kay}%
\ \BBA {} {Nieves-Chinchilla}%
}{%
{Kay}%
\ \BBA {} {Nieves-Chinchilla}%
}{%
{\protect \APACyear {2021}}%
}]{%
Kay2021}
\APACinsertmetastar {%
Kay2021}%
\begin{APACrefauthors}%
{Kay}, C.%
\BCBT {}\ \BBA {} {Nieves-Chinchilla}, T.%
\end{APACrefauthors}%
\unskip\
\newblock
\APACrefYearMonthDay{2021}{{\APACmonth{05}}}{}.
\newblock
{\BBOQ}\APACrefatitle {{Modeling Interplanetary Expansion and Deformation of
  CMEs With ANTEATR PARADE: 1. Relative Contribution of Different Forces}}
  {{Modeling Interplanetary Expansion and Deformation of CMEs With ANTEATR
  PARADE: 1. Relative Contribution of Different Forces}}.{\BBCQ}
\newblock
\APACjournalVolNumPages{Journal of Geophysical Research (Space
  Physics)}{126}{5}{e28911}.
\newblock
\begin{APACrefDOI} \doi{10.1029/2020JA028911} \end{APACrefDOI}
\PrintBackRefs{\CurrentBib}

\bibitem [\protect \citeauthoryear {%
{Kay}%
\ \BBA {} {Opher}%
}{%
{Kay}%
\ \BBA {} {Opher}%
}{%
{\protect \APACyear {2015}}%
}]{%
KayOpher2015}
\APACinsertmetastar {%
KayOpher2015}%
\begin{APACrefauthors}%
{Kay}, C.%
\BCBT {}\ \BBA {} {Opher}, M.%
\end{APACrefauthors}%
\unskip\
\newblock
\APACrefYearMonthDay{2015}{{\APACmonth{10}}}{}.
\newblock
{\BBOQ}\APACrefatitle {{The Heliocentric Distance where the Deflections and
  Rotations of Solar Coronal Mass Ejections Occur}} {{The Heliocentric Distance
  where the Deflections and Rotations of Solar Coronal Mass Ejections
  Occur}}.{\BBCQ}
\newblock
\APACjournalVolNumPages{\apjl}{811}{2}{L36}.
\newblock
\begin{APACrefDOI} \doi{10.1088/2041-8205/811/2/L36} \end{APACrefDOI}
\PrintBackRefs{\CurrentBib}

\bibitem [\protect \citeauthoryear {%
{Kilpua}%
, {Jian}%
, {Li}%
, {Luhmann}%
\BCBL {}\ \BBA {} {Russell}%
}{%
{Kilpua}%
\ \protect \BOthers {.}}{%
{\protect \APACyear {2012}}%
}]{%
Kilpua2012}
\APACinsertmetastar {%
Kilpua2012}%
\begin{APACrefauthors}%
{Kilpua}, E\BPBI K\BPBI J.%
, {Jian}, L\BPBI K.%
, {Li}, Y.%
, {Luhmann}, J\BPBI G.%
\BCBL {}\ \BBA {} {Russell}, C\BPBI T.%
\end{APACrefauthors}%
\unskip\
\newblock
\APACrefYearMonthDay{2012}{{\APACmonth{11}}}{}.
\newblock
{\BBOQ}\APACrefatitle {{Observations of ICMEs and ICME-like Solar Wind
  Structures from 2007 - 2010 Using Near-Earth and STEREO Observations}}
  {{Observations of ICMEs and ICME-like Solar Wind Structures from 2007 - 2010
  Using Near-Earth and STEREO Observations}}.{\BBCQ}
\newblock
\APACjournalVolNumPages{\solphys}{281}{1}{391-409}.
\newblock
\begin{APACrefDOI} \doi{10.1007/s11207-012-9957-0} \end{APACrefDOI}
\PrintBackRefs{\CurrentBib}

\bibitem [\protect \citeauthoryear {%
{Kilpua}%
, {Lugaz}%
, {Mays}%
\BCBL {}\ \BBA {} {Temmer}%
}{%
{Kilpua}%
\ \protect \BOthers {.}}{%
{\protect \APACyear {2019}}%
}]{%
Kilpua2019}
\APACinsertmetastar {%
Kilpua2019}%
\begin{APACrefauthors}%
{Kilpua}, E\BPBI K\BPBI J.%
, {Lugaz}, N.%
, {Mays}, M\BPBI L.%
\BCBL {}\ \BBA {} {Temmer}, M.%
\end{APACrefauthors}%
\unskip\
\newblock
\APACrefYearMonthDay{2019}{{\APACmonth{04}}}{}.
\newblock
{\BBOQ}\APACrefatitle {{Forecasting the Structure and Orientation of Earthbound
  Coronal Mass Ejections}} {{Forecasting the Structure and Orientation of
  Earthbound Coronal Mass Ejections}}.{\BBCQ}
\newblock
\APACjournalVolNumPages{Space Weather}{17}{4}{498-526}.
\newblock
\begin{APACrefDOI} \doi{10.1029/2018SW001944} \end{APACrefDOI}
\PrintBackRefs{\CurrentBib}

\bibitem [\protect \citeauthoryear {%
{Liu}%
\ \protect \BOthers {.}}{%
{Liu}%
\ \protect \BOthers {.}}{%
{\protect \APACyear {2014}}%
}]{%
Liu2014}
\APACinsertmetastar {%
Liu2014}%
\begin{APACrefauthors}%
{Liu}, Y\BPBI D.%
, {Luhmann}, J\BPBI G.%
, {Kajdi{\v{c}}}, P.%
, {Kilpua}, E\BPBI K\BPBI J.%
, {Lugaz}, N.%
, {Nitta}, N\BPBI V.%
\BDBL {}{Galvin}, A\BPBI B.%
\end{APACrefauthors}%
\unskip\
\newblock
\APACrefYearMonthDay{2014}{{\APACmonth{03}}}{}.
\newblock
{\BBOQ}\APACrefatitle {{Observations of an extreme storm in interplanetary
  space caused by successive coronal mass ejections}} {{Observations of an
  extreme storm in interplanetary space caused by successive coronal mass
  ejections}}.{\BBCQ}
\newblock
\APACjournalVolNumPages{Nature Communications}{5}{}{3481}.
\newblock
\begin{APACrefDOI} \doi{10.1038/ncomms4481} \end{APACrefDOI}
\PrintBackRefs{\CurrentBib}

\bibitem [\protect \citeauthoryear {%
{Lugaz}%
}{%
{Lugaz}%
}{%
{\protect \APACyear {2010}}%
}]{%
Lugaz2010}
\APACinsertmetastar {%
Lugaz2010}%
\begin{APACrefauthors}%
{Lugaz}, N.%
\end{APACrefauthors}%
\unskip\
\newblock
\APACrefYearMonthDay{2010}{{\APACmonth{12}}}{}.
\newblock
{\BBOQ}\APACrefatitle {{Accuracy and Limitations of Fitting and Stereoscopic
  Methods to Determine the Direction of Coronal Mass Ejections from
  Heliospheric Imagers Observations}} {{Accuracy and Limitations of Fitting and
  Stereoscopic Methods to Determine the Direction of Coronal Mass Ejections
  from Heliospheric Imagers Observations}}.{\BBCQ}
\newblock
\APACjournalVolNumPages{\solphys}{267}{2}{411-429}.
\newblock
\begin{APACrefDOI} \doi{10.1007/s11207-010-9654-9} \end{APACrefDOI}
\PrintBackRefs{\CurrentBib}

\bibitem [\protect \citeauthoryear {%
{Lugaz}%
\ \protect \BOthers {.}}{%
{Lugaz}%
\ \protect \BOthers {.}}{%
{\protect \APACyear {2012}}%
}]{%
Lugaz2012}
\APACinsertmetastar {%
Lugaz2012}%
\begin{APACrefauthors}%
{Lugaz}, N.%
, {Farrugia}, C\BPBI J.%
, {Davies}, J\BPBI A.%
, {M{\"o}stl}, C.%
, {Davis}, C\BPBI J.%
, {Roussev}, I\BPBI I.%
\BCBL {}\ \BBA {} {Temmer}, M.%
\end{APACrefauthors}%
\unskip\
\newblock
\APACrefYearMonthDay{2012}{{\APACmonth{11}}}{}.
\newblock
{\BBOQ}\APACrefatitle {{The Deflection of the Two Interacting Coronal Mass
  Ejections of 2010 May 23-24 as Revealed by Combined in Situ Measurements and
  Heliospheric Imaging}} {{The Deflection of the Two Interacting Coronal Mass
  Ejections of 2010 May 23-24 as Revealed by Combined in Situ Measurements and
  Heliospheric Imaging}}.{\BBCQ}
\newblock
\APACjournalVolNumPages{\apj}{759}{1}{68}.
\newblock
\begin{APACrefDOI} \doi{10.1088/0004-637X/759/1/68} \end{APACrefDOI}
\PrintBackRefs{\CurrentBib}

\bibitem [\protect \citeauthoryear {%
{Lugaz}%
\ \protect \BOthers {.}}{%
{Lugaz}%
\ \protect \BOthers {.}}{%
{\protect \APACyear {2010}}%
}]{%
LugazEtAl2010}
\APACinsertmetastar {%
LugazEtAl2010}%
\begin{APACrefauthors}%
{Lugaz}, N.%
, {Hernandez-Charpak}, J\BPBI N.%
, {Roussev}, I\BPBI I.%
, {Davis}, C\BPBI J.%
, {Vourlidas}, A.%
\BCBL {}\ \BBA {} {Davies}, J\BPBI A.%
\end{APACrefauthors}%
\unskip\
\newblock
\APACrefYearMonthDay{2010}{{\APACmonth{05}}}{}.
\newblock
{\BBOQ}\APACrefatitle {{Determining the Azimuthal Properties of Coronal Mass
  Ejections from Multi-Spacecraft Remote-Sensing Observations with STEREO
  SECCHI}} {{Determining the Azimuthal Properties of Coronal Mass Ejections
  from Multi-Spacecraft Remote-Sensing Observations with STEREO
  SECCHI}}.{\BBCQ}
\newblock
\APACjournalVolNumPages{\apj}{715}{1}{493-499}.
\newblock
\begin{APACrefDOI} \doi{10.1088/0004-637X/715/1/493} \end{APACrefDOI}
\PrintBackRefs{\CurrentBib}

\bibitem [\protect \citeauthoryear {%
{Manchester}%
\ \protect \BOthers {.}}{%
{Manchester}%
\ \protect \BOthers {.}}{%
{\protect \APACyear {2017}}%
}]{%
Manchester2017}
\APACinsertmetastar {%
Manchester2017}%
\begin{APACrefauthors}%
{Manchester}, W.%
, {Kilpua}, E\BPBI K\BPBI J.%
, {Liu}, Y\BPBI D.%
, {Lugaz}, N.%
, {Riley}, P.%
, {T{\"o}r{\"o}k}, T.%
\BCBL {}\ \BBA {} {Vr{\v{s}}nak}, B.%
\end{APACrefauthors}%
\unskip\
\newblock
\APACrefYearMonthDay{2017}{{\APACmonth{11}}}{}.
\newblock
{\BBOQ}\APACrefatitle {{The Physical Processes of CME/ICME Evolution}} {{The
  Physical Processes of CME/ICME Evolution}}.{\BBCQ}
\newblock
\APACjournalVolNumPages{\ssr}{212}{3-4}{1159-1219}.
\newblock
\begin{APACrefDOI} \doi{10.1007/s11214-017-0394-0} \end{APACrefDOI}
\PrintBackRefs{\CurrentBib}

\bibitem [\protect \citeauthoryear {%
{Manoharan}%
\ \protect \BOthers {.}}{%
{Manoharan}%
\ \protect \BOthers {.}}{%
{\protect \APACyear {2004}}%
}]{%
Manoharan2004}
\APACinsertmetastar {%
Manoharan2004}%
\begin{APACrefauthors}%
{Manoharan}, P\BPBI K.%
, {Gopalswamy}, N.%
, {Yashiro}, S.%
, {Lara}, A.%
, {Michalek}, G.%
\BCBL {}\ \BBA {} {Howard}, R\BPBI A.%
\end{APACrefauthors}%
\unskip\
\newblock
\APACrefYearMonthDay{2004}{{\APACmonth{06}}}{}.
\newblock
{\BBOQ}\APACrefatitle {{Influence of coronal mass ejection interaction on
  propagation of interplanetary shocks}} {{Influence of coronal mass ejection
  interaction on propagation of interplanetary shocks}}.{\BBCQ}
\newblock
\APACjournalVolNumPages{Journal of Geophysical Research (Space
  Physics)}{109}{A6}{A06109}.
\newblock
\begin{APACrefDOI} \doi{10.1029/2003JA010300} \end{APACrefDOI}
\PrintBackRefs{\CurrentBib}

\bibitem [\protect \citeauthoryear {%
{Manoharan}%
\ \BBA {} {Mujiber Rahman}%
}{%
{Manoharan}%
\ \BBA {} {Mujiber Rahman}%
}{%
{\protect \APACyear {2011}}%
}]{%
Manoharan2011}
\APACinsertmetastar {%
Manoharan2011}%
\begin{APACrefauthors}%
{Manoharan}, P\BPBI K.%
\BCBT {}\ \BBA {} {Mujiber Rahman}, A.%
\end{APACrefauthors}%
\unskip\
\newblock
\APACrefYearMonthDay{2011}{{\APACmonth{04}}}{}.
\newblock
{\BBOQ}\APACrefatitle {{Coronal mass ejections{\textemdash}Propagation time and
  associated internal energy}} {{Coronal mass ejections{\textemdash}Propagation
  time and associated internal energy}}.{\BBCQ}
\newblock
\APACjournalVolNumPages{Journal of Atmospheric and Solar-Terrestrial
  Physics}{73}{5-6}{671-677}.
\newblock
\begin{APACrefDOI} \doi{10.1016/j.jastp.2011.01.017} \end{APACrefDOI}
\PrintBackRefs{\CurrentBib}

\bibitem [\protect \citeauthoryear {%
{Mierla}%
\ \protect \BOthers {.}}{%
{Mierla}%
\ \protect \BOthers {.}}{%
{\protect \APACyear {2010}}%
}]{%
Mierla2010}
\APACinsertmetastar {%
Mierla2010}%
\begin{APACrefauthors}%
{Mierla}, M.%
, {Inhester}, B.%
, {Antunes}, A.%
, {Boursier}, Y.%
, {Byrne}, J\BPBI P.%
, {Colaninno}, R.%
\BDBL {}{Zhukov}, A\BPBI N.%
\end{APACrefauthors}%
\unskip\
\newblock
\APACrefYearMonthDay{2010}{{\APACmonth{01}}}{}.
\newblock
{\BBOQ}\APACrefatitle {{On the 3-D reconstruction of Coronal Mass Ejections
  using coronagraph data}} {{On the 3-D reconstruction of Coronal Mass
  Ejections using coronagraph data}}.{\BBCQ}
\newblock
\APACjournalVolNumPages{Annales Geophysicae}{28}{1}{203-215}.
\newblock
\begin{APACrefDOI} \doi{10.5194/angeo-28-203-2010} \end{APACrefDOI}
\PrintBackRefs{\CurrentBib}

\bibitem [\protect \citeauthoryear {%
{M{\"o}stl}%
\ \BBA {} {Davies}%
}{%
{M{\"o}stl}%
\ \BBA {} {Davies}%
}{%
{\protect \APACyear {2013}}%
}]{%
MoestlDavies2013}
\APACinsertmetastar {%
MoestlDavies2013}%
\begin{APACrefauthors}%
{M{\"o}stl}, C.%
\BCBT {}\ \BBA {} {Davies}, J\BPBI A.%
\end{APACrefauthors}%
\unskip\
\newblock
\APACrefYearMonthDay{2013}{{\APACmonth{07}}}{}.
\newblock
{\BBOQ}\APACrefatitle {{Speeds and Arrival Times of Solar Transients
  Approximated by Self-similar Expanding Circular Fronts}} {{Speeds and Arrival
  Times of Solar Transients Approximated by Self-similar Expanding Circular
  Fronts}}.{\BBCQ}
\newblock
\APACjournalVolNumPages{\solphys}{285}{1-2}{411-423}.
\newblock
\begin{APACrefDOI} \doi{10.1007/s11207-012-9978-8} \end{APACrefDOI}
\PrintBackRefs{\CurrentBib}

\bibitem [\protect \citeauthoryear {%
{M{\"o}stl}%
\ \protect \BOthers {.}}{%
{M{\"o}stl}%
\ \protect \BOthers {.}}{%
{\protect \APACyear {2015}}%
}]{%
Moestl2015}
\APACinsertmetastar {%
Moestl2015}%
\begin{APACrefauthors}%
{M{\"o}stl}, C.%
, {Rollett}, T.%
, {Frahm}, R\BPBI A.%
, {Liu}, Y\BPBI D.%
, {Long}, D\BPBI M.%
, {Colaninno}, R\BPBI C.%
\BDBL {}{Vr{\v{s}}nak}, B.%
\end{APACrefauthors}%
\unskip\
\newblock
\APACrefYearMonthDay{2015}{{\APACmonth{05}}}{}.
\newblock
{\BBOQ}\APACrefatitle {{Strong coronal channelling and interplanetary evolution
  of a solar storm up to Earth and Mars}} {{Strong coronal channelling and
  interplanetary evolution of a solar storm up to Earth and Mars}}.{\BBCQ}
\newblock
\APACjournalVolNumPages{Nature Communications}{6}{}{7135}.
\newblock
\begin{APACrefDOI} \doi{10.1038/ncomms8135} \end{APACrefDOI}
\PrintBackRefs{\CurrentBib}

\bibitem [\protect \citeauthoryear {%
{M{\"o}stl}%
\ \protect \BOthers {.}}{%
{M{\"o}stl}%
\ \protect \BOthers {.}}{%
{\protect \APACyear {2011}}%
}]{%
Moestl2011}
\APACinsertmetastar {%
Moestl2011}%
\begin{APACrefauthors}%
{M{\"o}stl}, C.%
, {Rollett}, T.%
, {Lugaz}, N.%
, {Farrugia}, C\BPBI J.%
, {Davies}, J\BPBI A.%
, {Temmer}, M.%
\BDBL {}{Biernat}, H\BPBI K.%
\end{APACrefauthors}%
\unskip\
\newblock
\APACrefYearMonthDay{2011}{{\APACmonth{11}}}{}.
\newblock
{\BBOQ}\APACrefatitle {{Arrival Time Calculation for Interplanetary Coronal
  Mass Ejections with Circular Fronts and Application to STEREO Observations of
  the 2009 February 13 Eruption}} {{Arrival Time Calculation for Interplanetary
  Coronal Mass Ejections with Circular Fronts and Application to STEREO
  Observations of the 2009 February 13 Eruption}}.{\BBCQ}
\newblock
\APACjournalVolNumPages{\apj}{741}{1}{34}.
\newblock
\begin{APACrefDOI} \doi{10.1088/0004-637X/741/1/34} \end{APACrefDOI}
\PrintBackRefs{\CurrentBib}

\bibitem [\protect \citeauthoryear {%
{M{\"o}stl}%
\ \protect \BOthers {.}}{%
{M{\"o}stl}%
\ \protect \BOthers {.}}{%
{\protect \APACyear {2020}}%
}]{%
Moestl2020}
\APACinsertmetastar {%
Moestl2020}%
\begin{APACrefauthors}%
{M{\"o}stl}, C.%
, {Weiss}, A\BPBI J.%
, {Bailey}, R\BPBI L.%
, {Reiss}, M\BPBI A.%
, {Amerstorfer}, U\BPBI V.%
, {Amerstorfer}, T.%
\BDBL {}{Stansby}, D.%
\end{APACrefauthors}%
\unskip\
\newblock
\APACrefYearMonthDay{2020}{{\APACmonth{07}}}{}.
\newblock
{\BBOQ}\APACrefatitle {{Prediction of the in situ coronal mass ejection rate
  for solar cycle 25: Implications for Parker Solar Probe in situ
  observations}} {{Prediction of the in situ coronal mass ejection rate for
  solar cycle 25: Implications for Parker Solar Probe in situ
  observations}}.{\BBCQ}
\newblock
\APACjournalVolNumPages{arXiv e-prints}{}{}{arXiv:2007.14743}.
\PrintBackRefs{\CurrentBib}

\bibitem [\protect \citeauthoryear {%
{Odstrcil}%
\ \protect \BOthers {.}}{%
{Odstrcil}%
\ \protect \BOthers {.}}{%
{\protect \APACyear {2004}}%
}]{%
Odstrcil2004}
\APACinsertmetastar {%
Odstrcil2004}%
\begin{APACrefauthors}%
{Odstrcil}, D.%
, {Pizzo}, V\BPBI J.%
, {Linker}, J\BPBI A.%
, {Riley}, P.%
, {Lionello}, R.%
\BCBL {}\ \BBA {} {Mikic}, Z.%
\end{APACrefauthors}%
\unskip\
\newblock
\APACrefYearMonthDay{2004}{{\APACmonth{10}}}{}.
\newblock
{\BBOQ}\APACrefatitle {{Initial coupling of coronal and heliospheric numerical
  magnetohydrodynamic codes}} {{Initial coupling of coronal and heliospheric
  numerical magnetohydrodynamic codes}}.{\BBCQ}
\newblock
\APACjournalVolNumPages{Journal of Atmospheric and Solar-Terrestrial
  Physics}{66}{15-16}{1311-1320}.
\newblock
\begin{APACrefDOI} \doi{10.1016/j.jastp.2004.04.007} \end{APACrefDOI}
\PrintBackRefs{\CurrentBib}

\bibitem [\protect \citeauthoryear {%
M.~{Owens}%
\ \protect \BOthers {.}}{%
M.~{Owens}%
\ \protect \BOthers {.}}{%
{\protect \APACyear {2020}}%
}]{%
Owens2020SoPh}
\APACinsertmetastar {%
Owens2020SoPh}%
\begin{APACrefauthors}%
{Owens}, M.%
, {Lang}, M.%
, {Barnard}, L.%
, {Riley}, P.%
, {Ben-Nun}, M.%
, {Scott}, C\BPBI J.%
\BDBL {}{Gonzi}, S.%
\end{APACrefauthors}%
\unskip\
\newblock
\APACrefYearMonthDay{2020}{{\APACmonth{03}}}{}.
\newblock
{\BBOQ}\APACrefatitle {{A Computationally Efficient, Time-Dependent Model of
  the Solar Wind for Use as a Surrogate to Three-Dimensional Numerical
  Magnetohydrodynamic Simulations}} {{A Computationally Efficient,
  Time-Dependent Model of the Solar Wind for Use as a Surrogate to
  Three-Dimensional Numerical Magnetohydrodynamic Simulations}}.{\BBCQ}
\newblock
\APACjournalVolNumPages{\solphys}{295}{3}{43}.
\newblock
\begin{APACrefDOI} \doi{10.1007/s11207-020-01605-3} \end{APACrefDOI}
\PrintBackRefs{\CurrentBib}

\bibitem [\protect \citeauthoryear {%
M\BPBI J.~{Owens}%
, {Lockwood}%
\BCBL {}\ \BBA {} {Barnard}%
}{%
M\BPBI J.~{Owens}%
\ \protect \BOthers {.}}{%
{\protect \APACyear {2017}}%
}]{%
Owens2017Nat}
\APACinsertmetastar {%
Owens2017Nat}%
\begin{APACrefauthors}%
{Owens}, M\BPBI J.%
, {Lockwood}, M.%
\BCBL {}\ \BBA {} {Barnard}, L\BPBI A.%
\end{APACrefauthors}%
\unskip\
\newblock
\APACrefYearMonthDay{2017}{{\APACmonth{06}}}{}.
\newblock
{\BBOQ}\APACrefatitle {{Coronal mass ejections are not coherent
  magnetohydrodynamic structures}} {{Coronal mass ejections are not coherent
  magnetohydrodynamic structures}}.{\BBCQ}
\newblock
\APACjournalVolNumPages{Scientific Reports}{7}{}{4152}.
\newblock
\begin{APACrefDOI} \doi{10.1038/s41598-017-04546-3} \end{APACrefDOI}
\PrintBackRefs{\CurrentBib}

\bibitem [\protect \citeauthoryear {%
{Paouris}%
\ \BBA {} {Mavromichalaki}%
}{%
{Paouris}%
\ \BBA {} {Mavromichalaki}%
}{%
{\protect \APACyear {2017}}%
}]{%
Paouris2017}
\APACinsertmetastar {%
Paouris2017}%
\begin{APACrefauthors}%
{Paouris}, E.%
\BCBT {}\ \BBA {} {Mavromichalaki}, H.%
\end{APACrefauthors}%
\unskip\
\newblock
\APACrefYearMonthDay{2017}{{\APACmonth{12}}}{}.
\newblock
{\BBOQ}\APACrefatitle {{Effective Acceleration Model for the Arrival Time of
  Interplanetary Shocks driven by Coronal Mass Ejections}} {{Effective
  Acceleration Model for the Arrival Time of Interplanetary Shocks driven by
  Coronal Mass Ejections}}.{\BBCQ}
\newblock
\APACjournalVolNumPages{\solphys}{292}{12}{180}.
\newblock
\begin{APACrefDOI} \doi{10.1007/s11207-017-1212-2} \end{APACrefDOI}
\PrintBackRefs{\CurrentBib}

\bibitem [\protect \citeauthoryear {%
{Pomoell}%
\ \BBA {} {Poedts}%
}{%
{Pomoell}%
\ \BBA {} {Poedts}%
}{%
{\protect \APACyear {2018}}%
}]{%
Pomoell2018}
\APACinsertmetastar {%
Pomoell2018}%
\begin{APACrefauthors}%
{Pomoell}, J.%
\BCBT {}\ \BBA {} {Poedts}, S.%
\end{APACrefauthors}%
\unskip\
\newblock
\APACrefYearMonthDay{2018}{{\APACmonth{06}}}{}.
\newblock
{\BBOQ}\APACrefatitle {{EUHFORIA: European heliospheric forecasting information
  asset}} {{EUHFORIA: European heliospheric forecasting information
  asset}}.{\BBCQ}
\newblock
\APACjournalVolNumPages{Journal of Space Weather and Space Climate}{8}{}{A35}.
\newblock
\begin{APACrefDOI} \doi{10.1051/swsc/2018020} \end{APACrefDOI}
\PrintBackRefs{\CurrentBib}

\bibitem [\protect \citeauthoryear {%
{Pulkkinen}%
}{%
{Pulkkinen}%
}{%
{\protect \APACyear {2007}}%
}]{%
Pulkkinen2007}
\APACinsertmetastar {%
Pulkkinen2007}%
\begin{APACrefauthors}%
{Pulkkinen}, T.%
\end{APACrefauthors}%
\unskip\
\newblock
\APACrefYearMonthDay{2007}{{\APACmonth{05}}}{}.
\newblock
{\BBOQ}\APACrefatitle {{Space Weather: Terrestrial Perspective}} {{Space
  Weather: Terrestrial Perspective}}.{\BBCQ}
\newblock
\APACjournalVolNumPages{Living Reviews in Solar Physics}{4}{1}{1}.
\newblock
\begin{APACrefDOI} \doi{10.12942/lrsp-2007-1} \end{APACrefDOI}
\PrintBackRefs{\CurrentBib}

\bibitem [\protect \citeauthoryear {%
{Reiss}%
\ \protect \BOthers {.}}{%
{Reiss}%
\ \protect \BOthers {.}}{%
{\protect \APACyear {2019}}%
}]{%
reiss19}
\APACinsertmetastar {%
reiss19}%
\begin{APACrefauthors}%
{Reiss}, M\BPBI A.%
, {MacNeice}, P\BPBI J.%
, {Mays}, L\BPBI M.%
, {Arge}, C\BPBI N.%
, {M{\"o}stl}, C.%
, {Nikolic}, L.%
\BCBL {}\ \BBA {} {Amerstorfer}, T.%
\end{APACrefauthors}%
\unskip\
\newblock
\APACrefYearMonthDay{2019}{{\APACmonth{02}}}{}.
\newblock
{\BBOQ}\APACrefatitle {{Forecasting the Ambient Solar Wind with Numerical
  Models. I. On the Implementation of an Operational Framework}} {{Forecasting
  the Ambient Solar Wind with Numerical Models. I. On the Implementation of an
  Operational Framework}}.{\BBCQ}
\newblock
\APACjournalVolNumPages{\apjs}{240}{}{35}.
\newblock
\begin{APACrefDOI} \doi{10.3847/1538-4365/aaf8b3} \end{APACrefDOI}
\PrintBackRefs{\CurrentBib}

\bibitem [\protect \citeauthoryear {%
{Reiss}%
\ \protect \BOthers {.}}{%
{Reiss}%
\ \protect \BOthers {.}}{%
{\protect \APACyear {2020}}%
}]{%
Reiss2020ApJ}
\APACinsertmetastar {%
Reiss2020ApJ}%
\begin{APACrefauthors}%
{Reiss}, M\BPBI A.%
, {MacNeice}, P\BPBI J.%
, {Muglach}, K.%
, {Arge}, C\BPBI N.%
, {M{\"o}stl}, C.%
, {Riley}, P.%
\BDBL {}{Amerstorfer}, U.%
\end{APACrefauthors}%
\unskip\
\newblock
\APACrefYearMonthDay{2020}{{\APACmonth{03}}}{}.
\newblock
{\BBOQ}\APACrefatitle {{Forecasting the Ambient Solar Wind with Numerical
  Models. II. An Adaptive Prediction System for Specifying Solar Wind Speed
  near the Sun}} {{Forecasting the Ambient Solar Wind with Numerical Models.
  II. An Adaptive Prediction System for Specifying Solar Wind Speed near the
  Sun}}.{\BBCQ}
\newblock
\APACjournalVolNumPages{\apj}{891}{2}{165}.
\newblock
\begin{APACrefDOI} \doi{10.3847/1538-4357/ab78a0} \end{APACrefDOI}
\PrintBackRefs{\CurrentBib}

\bibitem [\protect \citeauthoryear {%
{Richardson}%
\ \BBA {} {Cane}%
}{%
{Richardson}%
\ \BBA {} {Cane}%
}{%
{\protect \APACyear {2010}}%
}]{%
RichardsonCane2010}
\APACinsertmetastar {%
RichardsonCane2010}%
\begin{APACrefauthors}%
{Richardson}, I\BPBI G.%
\BCBT {}\ \BBA {} {Cane}, H\BPBI V.%
\end{APACrefauthors}%
\unskip\
\newblock
\APACrefYearMonthDay{2010}{Jun}{}.
\newblock
{\BBOQ}\APACrefatitle {{Near-Earth Interplanetary Coronal Mass Ejections During
  Solar Cycle 23 (1996 - 2009): Catalog and Summary of Properties}}
  {{Near-Earth Interplanetary Coronal Mass Ejections During Solar Cycle 23
  (1996 - 2009): Catalog and Summary of Properties}}.{\BBCQ}
\newblock
\APACjournalVolNumPages{\solphys}{264}{1}{189-237}.
\newblock
\begin{APACrefDOI} \doi{10.1007/s11207-010-9568-6} \end{APACrefDOI}
\PrintBackRefs{\CurrentBib}

\bibitem [\protect \citeauthoryear {%
{Richardson}%
\ \BBA {} {Cane}%
}{%
{Richardson}%
\ \BBA {} {Cane}%
}{%
{\protect \APACyear {2012}}%
}]{%
RichardsonCane2012}
\APACinsertmetastar {%
RichardsonCane2012}%
\begin{APACrefauthors}%
{Richardson}, I\BPBI G.%
\BCBT {}\ \BBA {} {Cane}, H\BPBI V.%
\end{APACrefauthors}%
\unskip\
\newblock
\APACrefYearMonthDay{2012}{{\APACmonth{05}}}{}.
\newblock
{\BBOQ}\APACrefatitle {{Near-earth solar wind flows and related geomagnetic
  activity during more than four solar cycles (1963-2011)}} {{Near-earth solar
  wind flows and related geomagnetic activity during more than four solar
  cycles (1963-2011)}}.{\BBCQ}
\newblock
\APACjournalVolNumPages{Journal of Space Weather and Space Climate}{2}{}{A02}.
\newblock
\begin{APACrefDOI} \doi{10.1051/swsc/2012003} \end{APACrefDOI}
\PrintBackRefs{\CurrentBib}

\bibitem [\protect \citeauthoryear {%
{Riley}%
\ \BBA {} {Lionello}%
}{%
{Riley}%
\ \BBA {} {Lionello}%
}{%
{\protect \APACyear {2011}}%
}]{%
riley11b}
\APACinsertmetastar {%
riley11b}%
\begin{APACrefauthors}%
{Riley}, P.%
\BCBT {}\ \BBA {} {Lionello}, R.%
\end{APACrefauthors}%
\unskip\
\newblock
\APACrefYearMonthDay{2011}{{\APACmonth{06}}}{}.
\newblock
{\BBOQ}\APACrefatitle {{Mapping Solar Wind Streams from the Sun to 1 AU: A
  Comparison of Techniques}} {{Mapping Solar Wind Streams from the Sun to 1 AU:
  A Comparison of Techniques}}.{\BBCQ}
\newblock
\APACjournalVolNumPages{\solphys}{270}{}{575-592}.
\newblock
\begin{APACrefDOI} \doi{10.1007/s11207-011-9766-x} \end{APACrefDOI}
\PrintBackRefs{\CurrentBib}

\bibitem [\protect \citeauthoryear {%
{Riley}%
\ \protect \BOthers {.}}{%
{Riley}%
\ \protect \BOthers {.}}{%
{\protect \APACyear {2018}}%
}]{%
Riley2018}
\APACinsertmetastar {%
Riley2018}%
\begin{APACrefauthors}%
{Riley}, P.%
, {Mays}, M\BPBI L.%
, {Andries}, J.%
, {Amerstorfer}, T.%
, {Biesecker}, D.%
, {Delouille}, V.%
\BDBL {}{Zhao}, X.%
\end{APACrefauthors}%
\unskip\
\newblock
\APACrefYearMonthDay{2018}{{\APACmonth{09}}}{}.
\newblock
{\BBOQ}\APACrefatitle {{Forecasting the Arrival Time of Coronal Mass Ejections:
  Analysis of the CCMC CME Scoreboard}} {{Forecasting the Arrival Time of
  Coronal Mass Ejections: Analysis of the CCMC CME Scoreboard}}.{\BBCQ}
\newblock
\APACjournalVolNumPages{Space Weather}{16}{9}{1245-1260}.
\newblock
\begin{APACrefDOI} \doi{10.1029/2018SW001962} \end{APACrefDOI}
\PrintBackRefs{\CurrentBib}

\bibitem [\protect \citeauthoryear {%
{Rollett}%
\ \protect \BOthers {.}}{%
{Rollett}%
\ \protect \BOthers {.}}{%
{\protect \APACyear {2016}}%
}]{%
Rollett2016}
\APACinsertmetastar {%
Rollett2016}%
\begin{APACrefauthors}%
{Rollett}, T.%
, {M{\"o}stl}, C.%
, {Isavnin}, A.%
, {Davies}, J\BPBI A.%
, {Kubicka}, M.%
, {Amerstorfer}, U\BPBI V.%
\BCBL {}\ \BBA {} {Harrison}, R\BPBI A.%
\end{APACrefauthors}%
\unskip\
\newblock
\APACrefYearMonthDay{2016}{Jun}{}.
\newblock
{\BBOQ}\APACrefatitle {{ElEvoHI: A Novel CME Prediction Tool for Heliospheric
  Imaging Combining an Elliptical Front with Drag-based Model Fitting}}
  {{ElEvoHI: A Novel CME Prediction Tool for Heliospheric Imaging Combining an
  Elliptical Front with Drag-based Model Fitting}}.{\BBCQ}
\newblock
\APACjournalVolNumPages{\apj}{824}{2}{131}.
\newblock
\begin{APACrefDOI} \doi{10.3847/0004-637X/824/2/131} \end{APACrefDOI}
\PrintBackRefs{\CurrentBib}

\bibitem [\protect \citeauthoryear {%
{Rollett}%
\ \protect \BOthers {.}}{%
{Rollett}%
\ \protect \BOthers {.}}{%
{\protect \APACyear {2014}}%
}]{%
Rollett2014}
\APACinsertmetastar {%
Rollett2014}%
\begin{APACrefauthors}%
{Rollett}, T.%
, {M{\"o}stl}, C.%
, {Temmer}, M.%
, {Frahm}, R\BPBI A.%
, {Davies}, J\BPBI A.%
, {Veronig}, A\BPBI M.%
\BDBL {}{Zhang}, T\BPBI L.%
\end{APACrefauthors}%
\unskip\
\newblock
\APACrefYearMonthDay{2014}{{\APACmonth{07}}}{}.
\newblock
{\BBOQ}\APACrefatitle {{Combined Multipoint Remote and in situ Observations of
  the Asymmetric Evolution of a Fast Solar Coronal Mass Ejection}} {{Combined
  Multipoint Remote and in situ Observations of the Asymmetric Evolution of a
  Fast Solar Coronal Mass Ejection}}.{\BBCQ}
\newblock
\APACjournalVolNumPages{\apjl}{790}{1}{L6}.
\newblock
\begin{APACrefDOI} \doi{10.1088/2041-8205/790/1/L6} \end{APACrefDOI}
\PrintBackRefs{\CurrentBib}

\bibitem [\protect \citeauthoryear {%
{Rouillard}%
\ \protect \BOthers {.}}{%
{Rouillard}%
\ \protect \BOthers {.}}{%
{\protect \APACyear {2008}}%
}]{%
Rouillard2008}
\APACinsertmetastar {%
Rouillard2008}%
\begin{APACrefauthors}%
{Rouillard}, A\BPBI P.%
, {Davies}, J\BPBI A.%
, {Forsyth}, R\BPBI J.%
, {Rees}, A.%
, {Davis}, C\BPBI J.%
, {Harrison}, R\BPBI A.%
\BDBL {}{Perry}, C\BPBI H.%
\end{APACrefauthors}%
\unskip\
\newblock
\APACrefYearMonthDay{2008}{{\APACmonth{05}}}{}.
\newblock
{\BBOQ}\APACrefatitle {{First imaging of corotating interaction regions using
  the STEREO spacecraft}} {{First imaging of corotating interaction regions
  using the STEREO spacecraft}}.{\BBCQ}
\newblock
\APACjournalVolNumPages{\grl}{35}{10}{L10110}.
\newblock
\begin{APACrefDOI} \doi{10.1029/2008GL033767} \end{APACrefDOI}
\PrintBackRefs{\CurrentBib}

\bibitem [\protect \citeauthoryear {%
{Ruffenach}%
\ \protect \BOthers {.}}{%
{Ruffenach}%
\ \protect \BOthers {.}}{%
{\protect \APACyear {2015}}%
}]{%
Ruffenach2015}
\APACinsertmetastar {%
Ruffenach2015}%
\begin{APACrefauthors}%
{Ruffenach}, A.%
, {Lavraud}, B.%
, {Farrugia}, C\BPBI J.%
, {D{\'e}moulin}, P.%
, {Dasso}, S.%
, {Owens}, M\BPBI J.%
\BDBL {}{Galvin}, A\BPBI B.%
\end{APACrefauthors}%
\unskip\
\newblock
\APACrefYearMonthDay{2015}{{\APACmonth{01}}}{}.
\newblock
{\BBOQ}\APACrefatitle {{Statistical study of magnetic cloud erosion by magnetic
  reconnection}} {{Statistical study of magnetic cloud erosion by magnetic
  reconnection}}.{\BBCQ}
\newblock
\APACjournalVolNumPages{Journal of Geophysical Research (Space
  Physics)}{120}{1}{43-60}.
\newblock
\begin{APACrefDOI} \doi{10.1002/2014JA020628} \end{APACrefDOI}
\PrintBackRefs{\CurrentBib}

\bibitem [\protect \citeauthoryear {%
{Savani}%
, {Owens}%
, {Rouillard}%
, {Forsyth}%
\BCBL {}\ \BBA {} {Davies}%
}{%
{Savani}%
\ \protect \BOthers {.}}{%
{\protect \APACyear {2010}}%
}]{%
Savani2010}
\APACinsertmetastar {%
Savani2010}%
\begin{APACrefauthors}%
{Savani}, N\BPBI P.%
, {Owens}, M\BPBI J.%
, {Rouillard}, A\BPBI P.%
, {Forsyth}, R\BPBI J.%
\BCBL {}\ \BBA {} {Davies}, J\BPBI A.%
\end{APACrefauthors}%
\unskip\
\newblock
\APACrefYearMonthDay{2010}{{\APACmonth{05}}}{}.
\newblock
{\BBOQ}\APACrefatitle {{Observational Evidence of a Coronal Mass Ejection
  Distortion Directly Attributable to a Structured Solar Wind}} {{Observational
  Evidence of a Coronal Mass Ejection Distortion Directly Attributable to a
  Structured Solar Wind}}.{\BBCQ}
\newblock
\APACjournalVolNumPages{\apjl}{714}{1}{L128-L132}.
\newblock
\begin{APACrefDOI} \doi{10.1088/2041-8205/714/1/L128} \end{APACrefDOI}
\PrintBackRefs{\CurrentBib}

\bibitem [\protect \citeauthoryear {%
{Schatten}%
}{%
{Schatten}%
}{%
{\protect \APACyear {1971}}%
}]{%
schatten71}
\APACinsertmetastar {%
schatten71}%
\begin{APACrefauthors}%
{Schatten}, K\BPBI H.%
\end{APACrefauthors}%
\unskip\
\newblock
\APACrefYearMonthDay{1971}{}{}.
\newblock
{\BBOQ}\APACrefatitle {{Current sheet magnetic model for the solar corona.}}
  {{Current sheet magnetic model for the solar corona.}}{\BBCQ}
\newblock
\APACjournalVolNumPages{Cosmic Electrodynamics}{2}{}{232-245}.
\PrintBackRefs{\CurrentBib}

\bibitem [\protect \citeauthoryear {%
{Schatten}%
, {Wilcox}%
\BCBL {}\ \BBA {} {Ness}%
}{%
{Schatten}%
\ \protect \BOthers {.}}{%
{\protect \APACyear {1969}}%
}]{%
schatten69}
\APACinsertmetastar {%
schatten69}%
\begin{APACrefauthors}%
{Schatten}, K\BPBI H.%
, {Wilcox}, J\BPBI M.%
\BCBL {}\ \BBA {} {Ness}, N\BPBI F.%
\end{APACrefauthors}%
\unskip\
\newblock
\APACrefYearMonthDay{1969}{{\APACmonth{03}}}{}.
\newblock
{\BBOQ}\APACrefatitle {{A model of interplanetary and coronal magnetic fields}}
  {{A model of interplanetary and coronal magnetic fields}}.{\BBCQ}
\newblock
\APACjournalVolNumPages{\solphys}{6}{}{442-455}.
\newblock
\begin{APACrefDOI} \doi{10.1007/BF00146478} \end{APACrefDOI}
\PrintBackRefs{\CurrentBib}

\bibitem [\protect \citeauthoryear {%
{Sheeley}%
, {Walters}%
, {Wang}%
\BCBL {}\ \BBA {} {Howard}%
}{%
{Sheeley}%
\ \protect \BOthers {.}}{%
{\protect \APACyear {1999}}%
}]{%
Sheeley1999}
\APACinsertmetastar {%
Sheeley1999}%
\begin{APACrefauthors}%
{Sheeley}, N\BPBI R.%
, {Walters}, J\BPBI H.%
, {Wang}, Y\BPBI M.%
\BCBL {}\ \BBA {} {Howard}, R\BPBI A.%
\end{APACrefauthors}%
\unskip\
\newblock
\APACrefYearMonthDay{1999}{{\APACmonth{11}}}{}.
\newblock
{\BBOQ}\APACrefatitle {{Continuous tracking of coronal outflows: Two kinds of
  coronal mass ejections}} {{Continuous tracking of coronal outflows: Two kinds
  of coronal mass ejections}}.{\BBCQ}
\newblock
\APACjournalVolNumPages{\jgr}{104}{A11}{24739-24768}.
\newblock
\begin{APACrefDOI} \doi{10.1029/1999JA900308} \end{APACrefDOI}
\PrintBackRefs{\CurrentBib}

\bibitem [\protect \citeauthoryear {%
{Shen}%
, {Wang}%
, {Gui}%
, {Ye}%
\BCBL {}\ \BBA {} {Wang}%
}{%
{Shen}%
\ \protect \BOthers {.}}{%
{\protect \APACyear {2011}}%
}]{%
Shen2011}
\APACinsertmetastar {%
Shen2011}%
\begin{APACrefauthors}%
{Shen}, C.%
, {Wang}, Y.%
, {Gui}, B.%
, {Ye}, P.%
\BCBL {}\ \BBA {} {Wang}, S.%
\end{APACrefauthors}%
\unskip\
\newblock
\APACrefYearMonthDay{2011}{{\APACmonth{04}}}{}.
\newblock
{\BBOQ}\APACrefatitle {{Kinematic Evolution of a Slow CME in Corona Viewed by
  STEREO-B on 8 October 2007}} {{Kinematic Evolution of a Slow CME in Corona
  Viewed by STEREO-B on 8 October 2007}}.{\BBCQ}
\newblock
\APACjournalVolNumPages{\solphys}{269}{2}{389-400}.
\newblock
\begin{APACrefDOI} \doi{10.1007/s11207-011-9715-8} \end{APACrefDOI}
\PrintBackRefs{\CurrentBib}

\bibitem [\protect \citeauthoryear {%
{Tappin}%
\ \BBA {} {Howard}%
}{%
{Tappin}%
\ \BBA {} {Howard}%
}{%
{\protect \APACyear {2009}}%
}]{%
TappinHoward2009_2}
\APACinsertmetastar {%
TappinHoward2009_2}%
\begin{APACrefauthors}%
{Tappin}, S\BPBI J.%
\BCBT {}\ \BBA {} {Howard}, T\BPBI A.%
\end{APACrefauthors}%
\unskip\
\newblock
\APACrefYearMonthDay{2009}{{\APACmonth{10}}}{}.
\newblock
{\BBOQ}\APACrefatitle {{Interplanetary Coronal Mass Ejections Observed in the
  Heliosphere: 2. Model and Data Comparison}} {{Interplanetary Coronal Mass
  Ejections Observed in the Heliosphere: 2. Model and Data Comparison}}.{\BBCQ}
\newblock
\APACjournalVolNumPages{\ssr}{147}{1-2}{55-87}.
\newblock
\begin{APACrefDOI} \doi{10.1007/s11214-009-9550-5} \end{APACrefDOI}
\PrintBackRefs{\CurrentBib}

\bibitem [\protect \citeauthoryear {%
{Temmer}%
, {Reiss}%
, {Nikolic}%
, {Hofmeister}%
\BCBL {}\ \BBA {} {Veronig}%
}{%
{Temmer}%
\ \protect \BOthers {.}}{%
{\protect \APACyear {2017}}%
}]{%
Temmer2017}
\APACinsertmetastar {%
Temmer2017}%
\begin{APACrefauthors}%
{Temmer}, M.%
, {Reiss}, M\BPBI A.%
, {Nikolic}, L.%
, {Hofmeister}, S\BPBI J.%
\BCBL {}\ \BBA {} {Veronig}, A\BPBI M.%
\end{APACrefauthors}%
\unskip\
\newblock
\APACrefYearMonthDay{2017}{{\APACmonth{02}}}{}.
\newblock
{\BBOQ}\APACrefatitle {{Preconditioning of Interplanetary Space Due to
  Transient CME Disturbances}} {{Preconditioning of Interplanetary Space Due to
  Transient CME Disturbances}}.{\BBCQ}
\newblock
\APACjournalVolNumPages{\apj}{835}{2}{141}.
\newblock
\begin{APACrefDOI} \doi{10.3847/1538-4357/835/2/141} \end{APACrefDOI}
\PrintBackRefs{\CurrentBib}

\bibitem [\protect \citeauthoryear {%
{Temmer}%
\ \protect \BOthers {.}}{%
{Temmer}%
\ \protect \BOthers {.}}{%
{\protect \APACyear {2011}}%
}]{%
temmer2011}
\APACinsertmetastar {%
temmer2011}%
\begin{APACrefauthors}%
{Temmer}, M.%
, {Rollett}, T.%
, {M{\"o}stl}, C.%
, {Veronig}, A\BPBI M.%
, {Vr{\v{s}}nak}, B.%
\BCBL {}\ \BBA {} {Odstr{\v{c}}il}, D.%
\end{APACrefauthors}%
\unskip\
\newblock
\APACrefYearMonthDay{2011}{{\APACmonth{12}}}{}.
\newblock
{\BBOQ}\APACrefatitle {{Influence of the Ambient Solar Wind Flow on the
  Propagation Behavior of Interplanetary Coronal Mass Ejections}} {{Influence
  of the Ambient Solar Wind Flow on the Propagation Behavior of Interplanetary
  Coronal Mass Ejections}}.{\BBCQ}
\newblock
\APACjournalVolNumPages{\apj}{743}{2}{101}.
\newblock
\begin{APACrefDOI} \doi{10.1088/0004-637X/743/2/101} \end{APACrefDOI}
\PrintBackRefs{\CurrentBib}

\bibitem [\protect \citeauthoryear {%
A.~{Thernisien}%
, {Vourlidas}%
\BCBL {}\ \BBA {} {Howard}%
}{%
A.~{Thernisien}%
\ \protect \BOthers {.}}{%
{\protect \APACyear {2009}}%
}]{%
Thernisien2009}
\APACinsertmetastar {%
Thernisien2009}%
\begin{APACrefauthors}%
{Thernisien}, A.%
, {Vourlidas}, A.%
\BCBL {}\ \BBA {} {Howard}, R\BPBI A.%
\end{APACrefauthors}%
\unskip\
\newblock
\APACrefYearMonthDay{2009}{May}{}.
\newblock
{\BBOQ}\APACrefatitle {{Forward Modeling of Coronal Mass Ejections Using
  STEREO/SECCHI Data}} {{Forward Modeling of Coronal Mass Ejections Using
  STEREO/SECCHI Data}}.{\BBCQ}
\newblock
\APACjournalVolNumPages{\solphys}{256}{1-2}{111-130}.
\newblock
\begin{APACrefDOI} \doi{10.1007/s11207-009-9346-5} \end{APACrefDOI}
\PrintBackRefs{\CurrentBib}

\bibitem [\protect \citeauthoryear {%
A\BPBI F\BPBI R.~{Thernisien}%
, {Howard}%
\BCBL {}\ \BBA {} {Vourlidas}%
}{%
A\BPBI F\BPBI R.~{Thernisien}%
\ \protect \BOthers {.}}{%
{\protect \APACyear {2006}}%
}]{%
Thernisien2006}
\APACinsertmetastar {%
Thernisien2006}%
\begin{APACrefauthors}%
{Thernisien}, A\BPBI F\BPBI R.%
, {Howard}, R\BPBI A.%
\BCBL {}\ \BBA {} {Vourlidas}, A.%
\end{APACrefauthors}%
\unskip\
\newblock
\APACrefYearMonthDay{2006}{Nov}{}.
\newblock
{\BBOQ}\APACrefatitle {{Modeling of Flux Rope Coronal Mass Ejections}}
  {{Modeling of Flux Rope Coronal Mass Ejections}}.{\BBCQ}
\newblock
\APACjournalVolNumPages{\apj}{652}{1}{763-773}.
\newblock
\begin{APACrefDOI} \doi{10.1086/508254} \end{APACrefDOI}
\PrintBackRefs{\CurrentBib}

\bibitem [\protect \citeauthoryear {%
{Vr{\v{s}}nak}%
\ \protect \BOthers {.}}{%
{Vr{\v{s}}nak}%
\ \protect \BOthers {.}}{%
{\protect \APACyear {2013}}%
}]{%
Vrsnak2013}
\APACinsertmetastar {%
Vrsnak2013}%
\begin{APACrefauthors}%
{Vr{\v{s}}nak}, B.%
, {{\v{Z}}ic}, T.%
, {Vrbanec}, D.%
, {Temmer}, M.%
, {Rollett}, T.%
, {M{\"o}stl}, C.%
\BDBL {}{Shanmugaraju}, A.%
\end{APACrefauthors}%
\unskip\
\newblock
\APACrefYearMonthDay{2013}{{\APACmonth{07}}}{}.
\newblock
{\BBOQ}\APACrefatitle {{Propagation of Interplanetary Coronal Mass Ejections:
  The Drag-Based Model}} {{Propagation of Interplanetary Coronal Mass
  Ejections: The Drag-Based Model}}.{\BBCQ}
\newblock
\APACjournalVolNumPages{\solphys}{285}{1-2}{295-315}.
\newblock
\begin{APACrefDOI} \doi{10.1007/s11207-012-0035-4} \end{APACrefDOI}
\PrintBackRefs{\CurrentBib}

\bibitem [\protect \citeauthoryear {%
{{\v{Z}}ic}%
, {Vr{\v{s}}nak}%
\BCBL {}\ \BBA {} {Temmer}%
}{%
{{\v{Z}}ic}%
\ \protect \BOthers {.}}{%
{\protect \APACyear {2015}}%
}]{%
Zic2015}
\APACinsertmetastar {%
Zic2015}%
\begin{APACrefauthors}%
{{\v{Z}}ic}, T.%
, {Vr{\v{s}}nak}, B.%
\BCBL {}\ \BBA {} {Temmer}, M.%
\end{APACrefauthors}%
\unskip\
\newblock
\APACrefYearMonthDay{2015}{{\APACmonth{06}}}{}.
\newblock
{\BBOQ}\APACrefatitle {{Heliospheric Propagation of Coronal Mass Ejections:
  Drag-based Model Fitting}} {{Heliospheric Propagation of Coronal Mass
  Ejections: Drag-based Model Fitting}}.{\BBCQ}
\newblock
\APACjournalVolNumPages{\apjs}{218}{2}{32}.
\newblock
\begin{APACrefDOI} \doi{10.1088/0067-0049/218/2/32} \end{APACrefDOI}
\PrintBackRefs{\CurrentBib}

\bibitem [\protect \citeauthoryear {%
Y.~{Wang}%
\ \protect \BOthers {.}}{%
Y.~{Wang}%
\ \protect \BOthers {.}}{%
{\protect \APACyear {2016}}%
}]{%
Wang2016}
\APACinsertmetastar {%
Wang2016}%
\begin{APACrefauthors}%
{Wang}, Y.%
, {Zhang}, Q.%
, {Liu}, J.%
, {Shen}, C.%
, {Shen}, F.%
, {Yang}, Z.%
\BDBL {}{Zhuang}, B.%
\end{APACrefauthors}%
\unskip\
\newblock
\APACrefYearMonthDay{2016}{{\APACmonth{08}}}{}.
\newblock
{\BBOQ}\APACrefatitle {{On the propagation of a geoeffective coronal mass
  ejection during 15-17 March 2015}} {{On the propagation of a geoeffective
  coronal mass ejection during 15-17 March 2015}}.{\BBCQ}
\newblock
\APACjournalVolNumPages{Journal of Geophysical Research (Space
  Physics)}{121}{8}{7423-7434}.
\newblock
\begin{APACrefDOI} \doi{10.1002/2016JA022924} \end{APACrefDOI}
\PrintBackRefs{\CurrentBib}

\bibitem [\protect \citeauthoryear {%
Y\BHBI M.~{Wang}%
\ \BBA {} {Sheeley}%
}{%
Y\BHBI M.~{Wang}%
\ \BBA {} {Sheeley}%
}{%
{\protect \APACyear {1995}}%
}]{%
wang95}
\APACinsertmetastar {%
wang95}%
\begin{APACrefauthors}%
{Wang}, Y\BHBI M.%
\BCBT {}\ \BBA {} {Sheeley}, N\BPBI R., Jr.%
\end{APACrefauthors}%
\unskip\
\newblock
\APACrefYearMonthDay{1995}{{\APACmonth{07}}}{}.
\newblock
{\BBOQ}\APACrefatitle {{Solar Implications of ULYSSES Interplanetary Field
  Measurements}} {{Solar Implications of ULYSSES Interplanetary Field
  Measurements}}.{\BBCQ}
\newblock
\APACjournalVolNumPages{\apjl}{447}{}{L143}.
\newblock
\begin{APACrefDOI} \doi{10.1086/309578} \end{APACrefDOI}
\PrintBackRefs{\CurrentBib}

\bibitem [\protect \citeauthoryear {%
{Zhuang}%
\ \protect \BOthers {.}}{%
{Zhuang}%
\ \protect \BOthers {.}}{%
{\protect \APACyear {2017}}%
}]{%
Zhuang2017}
\APACinsertmetastar {%
Zhuang2017}%
\begin{APACrefauthors}%
{Zhuang}, B.%
, {Wang}, Y.%
, {Shen}, C.%
, {Liu}, S.%
, {Wang}, J.%
, {Pan}, Z.%
\BDBL {}{Liu}, R.%
\end{APACrefauthors}%
\unskip\
\newblock
\APACrefYearMonthDay{2017}{{\APACmonth{08}}}{}.
\newblock
{\BBOQ}\APACrefatitle {{The Significance of the Influence of the CME Deflection
  in Interplanetary Space on the CME Arrival at Earth}} {{The Significance of
  the Influence of the CME Deflection in Interplanetary Space on the CME
  Arrival at Earth}}.{\BBCQ}
\newblock
\APACjournalVolNumPages{\apj}{845}{2}{117}.
\newblock
\begin{APACrefDOI} \doi{10.3847/1538-4357/aa7fc0} \end{APACrefDOI}
\PrintBackRefs{\CurrentBib}

\bibitem [\protect \citeauthoryear {%
{Zuccarello}%
\ \protect \BOthers {.}}{%
{Zuccarello}%
\ \protect \BOthers {.}}{%
{\protect \APACyear {2012}}%
}]{%
Zuccarello2012}
\APACinsertmetastar {%
Zuccarello2012}%
\begin{APACrefauthors}%
{Zuccarello}, F\BPBI P.%
, {Bemporad}, A.%
, {Jacobs}, C.%
, {Mierla}, M.%
, {Poedts}, S.%
\BCBL {}\ \BBA {} {Zuccarello}, F.%
\end{APACrefauthors}%
\unskip\
\newblock
\APACrefYearMonthDay{2012}{{\APACmonth{01}}}{}.
\newblock
{\BBOQ}\APACrefatitle {{The Role of Streamers in the Deflection of Coronal Mass
  Ejections: Comparison between STEREO Three-dimensional Reconstructions and
  Numerical Simulations}} {{The Role of Streamers in the Deflection of Coronal
  Mass Ejections: Comparison between STEREO Three-dimensional Reconstructions
  and Numerical Simulations}}.{\BBCQ}
\newblock
\APACjournalVolNumPages{\apj}{744}{1}{66}.
\newblock
\begin{APACrefDOI} \doi{10.1088/0004-637X/744/1/66} \end{APACrefDOI}
\PrintBackRefs{\CurrentBib}

\end{thebibliography}




\end{document}